\documentstyle[12pt,epsfig]{article}

\def\be{\begin{equation}}
\def\ee{\end{equation}}
\def\ba{\begin{eqnarray}}
\def\ea{\end{eqnarray}}
\def\ga{\mathrel{\raise.3ex\hbox{$>$\kern-.75em\lower1ex\hbox{$\sim$}}}}
\def\la{\mathrel{\raise.3ex\hbox{$<$\kern-.75em\lower1ex\hbox{$\sim$}}}}

\newcommand{\daa}{\Delta\alpha/\alpha}

\newcommand{\bi}[1]{\bibitem{#1}}
\newcommand{\fr}[2]{\frac{#1}{#2}}

\begin{document}

\baselineskip=16pt
\begin{titlepage}
\rightline{UMN--TH--2313/04}
\rightline{FTPI--MINN--04/24}
\rightline{UVIC--TH--04/06}
\rightline{hep-ph/0406039}
\rightline{June 2004}
\begin{center}

\vspace{0.5cm}

\large {\bf Quintessence Models and the Cosmological Evolution of $\alpha$.}
\vspace*{5mm}
\normalsize

{\bf Seokcheon Lee$^{\,(a)}$},  {\bf Keith A. Olive$^{\, (a,b)}$ and
{\bf Maxim Pospelov$^{\,(c)}$}}

\smallskip
\medskip

$^{(a)}${\it School of Physics and Astronomy,\\ University of Minnesota,
 Minneapolis, MN 55455, USA}

$^{(b)}${\it William I. Fine Theoretical Physics Institute,\\  University of
Minnesota, Minneapolis, MN 55455, USA}

$^c${\it Department of Physics and Astronomy, \\ University of Victoria,
Victoria, BC, V8P 1A1 Canada}

\smallskip
\end{center}
\vskip0.6in

\centerline{\large\bf Abstract}
The cosmological evolution of a quintessence-like scalar field $\phi$
coupled to matter and gauge fields leads to effective
modifications of the coupling constants and particle masses over time.
We analyze a class of models where the scalar field potential
$V(\phi)$ and the couplings to matter $B(\phi)$ admit common extremum in
$\phi$, as in the Damour-Polyakov ansatz. We find that even for the simplest 
choices of potentials and $B(\phi)$, the observational constraints on 
$\daa$ coming from quasar absorption spectra, the Oklo phenomenon and
Big Bang nucleosynthesis provide complementary constraints on the 
parameters of the model. We show the evolutionary history of these models in
some detail and describe the effects of a varying mass for dark matter.

\vspace*{2mm}

\end{titlepage}

\section{Introduction}
\setcounter{equation}{0}

The existence of dark energy supported by a number of
cosmological observations remains one of the most outstanding
puzzles in modern cosmology \cite{SCP}.
The cosmological
constant and/or a quintessence field are the most commonly
proposed candidates for dark energy.
The possibility that a scalar field at early cosmological
times follows an attractor-type solution \cite{Ratra,FJ} and
tracks the evolution of the visible matter-energy density
while dominating the energy density in the Universe at late cosmological
stages has been
a subject of lively debates in the cosmology literature \cite{Steinhardt} - \cite{BK}.
There are certain hopes that quintessence-like models
may help alleviate the severe fine-tuning associated with
the cosmological constant problem. This by itself represents a
legitimate field of research and has triggered
various model building efforts.

The most fundamental difference between quintessence and the cosmological
constant is that the former represents a new very light degree of freedom,
with a ``wavelength'' comparable to the size of the Hubble horizon.
As a consequence, the dynamical scalar field gives rise to a time-dependent
equation of state parameter,
$\omega_{\phi} \equiv p_{\phi} / \rho_{\phi}$.
Therefore, the variation of $\omega$ for dark energy with
redshift is among the most important cosmological parameters
to measure, and will be a large component of future efforts in cosmology.

Are there additional experimental ways of checking for the existence or absence of
dark energy in the form of quintessence? Apart from ``geometric''
tests using standard cosmological methods (CMB, supernovae, lensing, etc.),
one could also hope to detect the possible interaction of the quintessence
field with matter. This can occur either through the observation of the
effective change of the coupling constants and masses of particles over
cosmological times, or via detecting additional components to the gravitational
interaction due to exchange by the scalar. A positive detection of the
cosmological change of coupling constants would be firm proof of the existence of
new degree(s) of freedom  with extremely long wavelength,
thus providing a perfect candidate for quintessence.

For this reason, the recently reported indications of a change in the effective
electromagnetic coupling constant \cite{Webb}, $\Delta \alpha / \alpha \sim -0.6 \times 10^{-5}$
 at redshifts $z\sim 1$, if indeed true, can be
considered as an independent detection of an ultra-light degree of freedom.
These results \cite{Webb} are based on the comparison of the relativistic shifts of atomic energy
levels in quasar
absorption spectra. For some time this result remained unchallenged, yet recently there
have been attempts to detect a variation in $\alpha$ using similar methods \cite{Petitjean}
that has so far  led to null results. Also, alternative ($\alpha$-unrelated)
explanations of the results of Ref.~\cite{Webb} based
on a $z$-dependent Mg isotope abundance were shown to be astrophysically viable \cite{AMO}.
This rather controversial situation becomes even more complicated if other
constraints on $\Delta \alpha / \alpha $ are taken into account \cite{Uzan}. The Oklo natural
reactor and meteoritic abundances of rhenium provide stringent constraints on
the change of the coupling constants that goes back to $z\sim 0.1 - 0.4$ \cite{Oklo,OPQ,fuj}.
However, a recent re-analysis of Oklo phenomenon suggested that the present data are consistent
with the non-zero change of $\Delta\alpha/\alpha = 4.5 \times 10^{-8}$  \cite{Lam}.

Despite the questionable status of the non-zero claim for
$\Delta \alpha/\alpha$ there have been a number of attempts \cite{Chiba} - \cite{CNP},  \cite{Bek} - \cite{OP}
to build simple models that could account for a possible $O(10^{-5})$
relative shift in the fine structure constant at redshifts $z\sim 1$.
It is widely accepted now that the minimal Bekenstein model \cite{Bek} with a
scalar field coupled {\em only} to the electromagnetic field
in the initial Lagrangian,
\be
{\cal L}_{\varphi F} = -\fr{1}{4}B_F\left(\fr{\varphi}{M_{\rm Pl}}\right) F_{\mu\nu}F^{\mu \nu }
= -\fr{1}{4}\left(1+\zeta_F\fr{\varphi}{M_{\rm Pl}} +\fr{\xi_F}{2}
\left(\fr{\varphi}{M_{\rm Pl}}\right)^2+...\right)
F_{\mu\nu}F^{\mu \nu },
\label{intro}
\ee
cannot provide $\Delta \alpha/\alpha$
larger than $10^{-9}-10^{-10}$ level when the constraints on the nonuniversal
gravitational interaction mediated by this field are imposed.
This is because the electromagnetic portion of the baryon energy density
that drives the cosmological evolution of $\varphi$ is a very weak source.
It has been suggested that the coupling to dark matter \cite{Damour,OP,SBM}
and/or the self-interaction potential \cite{DZ}
of the scalar field drives its evolution. A number of tracker field and
quintessence field models coupled to $F_{\mu\nu}^2$ have been analyzed with the
universal conclusion that a linear coupling to the electromagnetic energy density
on the order of $10^{-3}\la \zeta_F\la 10^{-5}$ can easily explain the QSO-suggested change of
the fine structure constant and satisfy the experimental limits on the
universality of the gravitational interaction. It has proven to be a harder
task to have a substantial change in the fine structure constant at $z\sim 1$ and
remain consistent with the Oklo and meteoritic constraints. Only a few models (see {\em e.g.}
Refs. \cite{Wett}, \cite{CNP}) are known to pass these requirements.

To this end, it will be useful to investigate a class of quintessence models where the
potential $V(\varphi)$ admits a minimum at $\varphi_0$, and $\varphi(z)$ approaches
$\varphi_0$ when $z\to 0$. Without  loss of generality, we can put the extremum
value to zero, and then $V(0)\simeq  \Lambda = \Omega_\Lambda\rho_c$, where $\rho_c$ is
the closure energy density today. In order to suppress the variation of $\alpha$ at late
times, one could also assume that $\zeta_F=0$, or in other words, require
$V(\varphi)$ and $ B_F(\varphi)$ to share the same extremum.
We note that most of the quintessence models analyzed in the
context of changing coupling constants to date, were coupled to the electromagnetic
lagrangian linearly, while the coupling to other gauge and matter
fields was neglected. Continuing along the same line, it would be reasonable
to allow a multitude of couplings $B_i(\varphi)$, and require $\zeta_i =0$.
This is exactly the proposal of the ``least coupling" principle, introduced
a decade ago by Damour and Nordtvedt \cite{DN} and 
Damour and Polyakov \cite{DP} in the context of the
string dilaton with the primary goal to suppress the strength of the gravitational
interaction mediated by $\varphi$. Here we supplement their idea by including
the self-interaction potential $V(\varphi)$ with an extremum at the same value of  $\varphi$ as in $B_i(\varphi)$.
We note that this property, a common extremum in $\varphi$, remains intact even in the
presence of the radiative corrections, and therefore does not require any additional
fine-tuning.

In what follows, we analyze in detail the cosmology of  coupled quintessence models 
with a common extremum in $V(\varphi)$ and gauge and matter/gauge couplings $B_i(\varphi)$.
We note that only a coupled quintessence model derived from
Brans-Dicke theory has been analyzed to date \cite{Amendola,Wands}.
We investigate the size of the possible variation of masses and couplings in
these models and find that some generic choices of $V(\varphi)$ and $B_i(\varphi)$ can be
consistent with all observational requirements. This paper is organized as follows.
In the next section we introduce our model and display the necessary cosmological equations.
In sections 3-5, we make specific choices for the potential $V(\varphi)$ and matter/gauge
couplings $B_i(\varphi)$ and find the evolution of the dark energy and dark matter energy density
over the redshift. We make predictions for the variation of the coupling constants and
masses and outline the choice of parameters that satisfies all existing constraints.
We reach our conclusions in section 6.

\section{Cosmological evolution of a scalar field in the presence of couplings to matter}
\setcounter{equation}{0}
We start by writing down the general equation for the interaction
of a light scalar field $\varphi$ with matter,
\begin{eqnarray}
S_{\phi} = \int d^4x \sqrt{-g} \Biggl\{ \frac{\bar{M}^2}{2}
[\partial^{\mu} \phi
\partial_{\mu} \phi - R] - V(\phi) \nonumber \\
- \frac{B_{Fi}(\phi)}{4}F^{(i)}_{\mu\nu}F^{(i)\mu\nu} + \sum_j
[\bar\psi_j iD\!\!\!\!/ \psi_j -
B_j(\phi)m_j\bar\psi_j\psi_j]\Biggr\}. \label{lagrangian}
\end{eqnarray}
In this expression, $\phi$ is the scalar field of interest with
its kinetic term normalized together with the gravitational
interaction. The placement of the reduced Planck mass, $\bar M$, in (\ref{lagrangian})
renders $\phi$ dimensionless. $B_{Fi}(\phi)$ represents the
$\phi$-dependence of  the gauge couplings in Standard Model where 
the sum is over all three groups. $\psi_j$ represents Standard Model
fermions that are coupled to $\phi$ via the functions $B_j(\phi)$.
Note that the $\phi$-dependent rescaling of the matter fields and
metric allows us to remove $\phi$ from the couplings to the kinetic
terms and $R$. A similar rescaling in the gauge sector will lead to
the appearance of a $\phi$-dependence in the interaction terms
between gauge fields and matter, {\em i.e.} precisely 
$\phi$-dependent gauge couplings. Besides SM fermions, the
sum over $j$ includes other matter field ({\em i.e.} scalar
Higgses, Majorana neutrinos etc.) as well as various interaction
terms. Finally, any viable cosmological scenario requires cold
dark matter, and we also include it in the sum with the separate
coupling $B_{DM}(\phi)$. Since $\phi$ couples to the trace $T_\mu^\mu$
of dark matter, our results are independent of  the nature of the dark matter particles (scalars or
fermions). Furthermore,  since dark matter is the dominant component of the
matter energy-density, we can simply take $B_m(\phi) \simeq
B_{DM}(\phi)$.

{}From the above action (\ref{lagrangian}) we can derive the
equation of motion for the scalar field, $\phi$: 
\be 
\Box \phi +
\frac{1}{\bar{M}^2} \frac{\partial V}{\partial \phi} = - \sum_{j}
\frac{1}{\bar{M}^2} \frac{\partial B_{j}}{\partial \phi} m_{j}
\langle \bar{\psi}_{j} \psi_{j} \rangle ,
\label{fieldeq} 
\ee 
where $\langle \bar{\psi}_{j} \psi_{j} \rangle $ stands for the number density of 
a $j$-th fermion. Notice that at tree level, radiation does not contribute to the 
r.h.s. of Eq. (\ref{fieldeq}) because its stress-energy tensor is 
traceless.
Given a potential $V(\phi)$, the perfect fluid energy density and
pressure contributions due to $\phi$ are: 
\ba
\rho_{\phi} &=& \frac{1}{2} \bar{M}^2 \dot{\phi}^2 + V(\phi) \label{rho} \\
p_{\phi} &=& \frac{1}{2} \bar{M}^2 \dot{\phi}^2 - V(\phi) \equiv
\omega_{\phi} \rho_{\phi} \label{p}, 
\ea
where the parameter $\omega_\phi$ is related to the equation of
state (EOS) of the scalar field. Combining  Eqs. (\ref{rho}) and
(\ref{p}), we obtain a useful relation between the potential $V(\phi)$, 
the energy density of 
the scalar field, and $\omega_\phi$,
 \be \rho_{\phi} = \frac{2
V(\phi)}{(1 - \omega_{\phi})}. 
\label{rho1} 
\ee 
The energy density $\rho_\phi$ and 
pressure $p_\phi$ of the scalar field contribute to
the r.h.s. of Einstein's equations and yield 
the following Friedmann equation in a Robertson-Walker Universe, 
\ba
H^2 &=& \frac{1}{3 \bar{M}^2} \Bigl( \rho_{\phi} + \rho_{r} +
\rho_{m} \Bigr) \equiv \frac{1}{3 \bar{M}^2} \rho_{cr} 
\label{H},
\ea 
where $H=\dot a/a $ is the Hubble expansion rate,
$\bar{M}^2 = M_{p}^2 / 8 \pi$ is a reduced Planck mass,
$\rho_{r}$ and $\rho_{m}$ is the energy density of radiation and
matter, and $\rho_{cr}$ is the critical energy density. The energy density 
of matter and radiation contains $\phi$-dependence, {\em i.e.} $\rho_{m} =
\sum_{j}B_{j}(\phi)m_{j}\langle \bar{\psi}_j\psi_j\rangle$.
Using these definitions, we can rewrite the scalar field equation (\ref{fieldeq}), 
\be 
\ddot{\phi} +
3 H \dot{\phi} + \frac{1}{\bar{M}^2} \frac{\partial V}{\partial
\phi} = - \frac{1}{\bar{M}^2} \frac{\partial \ln B_{m}}{\partial
\phi} \rho_{m} .
\label{eqphi} 
\ee 
Conservation of the energy-momentum tensor gives us another equation,
\ba
\dot{\rho}_{\phi} + 3H (1 + \omega_{\phi}) \rho_{\phi} &=& -
\frac{\partial \ln B_{m}}{\partial \phi} \rho_{m} \dot{\phi}.
\ea

For cosmological studies that span a large range of redshifts $z$,
it is convenient to introduce the variable $x$ as the logarithm of the scale factor $a$,
\be 
x= \ln a = - \ln (1 + z) 
\label{x} 
\ee 
where we choose the present scale factor $a^{(0)} = 1$.  With the use of the 
variable $x$, we can rewrite the relevant system of equations for $d\ln \rho_i/dx$ in the
following form,
\ba 
\frac{d \ln \rho_{m}}{d x} &=& - 3 ( 1 + \omega_{m} ) +
\frac{\partial \ln B_{m}(\phi)}{\partial \phi} \frac{d \phi}{d x},
\label{rhoi1} \\ 
\frac{d \ln \rho_{r}}{d x} &=& - 3 ( 1 +
\omega_{r} ), 
\label{rhor1} 
\\
\frac{d \ln \rho_{\phi}}{d
x } &=& -3 ( 1 + \omega_{\phi} ) - \frac{\partial \ln
B_{m}(\phi)}{\partial \phi} \frac{d \phi}{d x}
\frac{\rho_{m}}{\rho_{\phi}},
\label{rhophi1} 
\ea 
where $\omega_{r} = 1/3$ and $\omega_{m} = 0$ should be used for radiation and matter 
respectively. 

The evolution of $\phi$ as a function of  redshift
can be found relatively easy if one is able  to solve the
cosmological equations and obtain $\omega_\phi$ and $\Omega_i \equiv \rho_i/\rho_{cr}$
as a function of $x$. For example, the  derivative of 
$\phi$ with respect to $x$ can be obtained
from  Eq (\ref{rho}),
\be 
\left(\frac{d \phi}{d x}\right)^2 = 3
\Omega_{\phi} ( 1 + \omega_{\phi} ),
\label{dphi} 
\ee 
If $\omega_{\phi}$ is close to $-1$, the kinetic energy of the scalar
field goes to zero which occurs when the scalar field is
close to the minimum of the potential.

The evolution of $\phi(x)$ expressed in terms of 
$\omega_\phi$ and $\Omega_\phi$ takes an especially simple 
form when $B_m(\phi)$ does not produce a significant change in the cosmological 
equations, and terms containing $\partial B_m(\phi)/\partial \phi$ can be neglected.
In this case, the matter and radiation energy densities can be easily related, 
$\rho_{r} = (a_{eq}/a) \rho_{m}$, where $a_{eq}$ is the scale factor when 
the radiation and  matter densities are equal, and $a_{eq}\ll 1$. This allows 
us to rewrite Eq. (\ref{H}) for the evolution of the scalar field energy density 
in a useful form,
\be 
\frac{d \ln(1 -\Omega_{\phi})}{d x} =  \Omega_{\phi}
\Biggl[3 \omega_{\phi} 
- \Bigl(\frac{a_{eq}}{a+ a_{eq}} \Bigr)\
\Biggr].
\label{dOmega} 
\ee 
The dependence of critical density on the scale factor can be found explicitly,
which allows us to express $V(\phi)$ from (\ref{rho1}) as
\be
V(\phi)= \fr 12 (1-\omega_\phi)\Omega_\phi\rho_{cr}= \fr 32 (\bar MH^{(0)})^2
(1-\omega_\phi)\Omega_\phi\fr{1-\Omega^{(0)}_\phi}{1-\Omega_\phi}\left(\fr{a^{(0)}}{a}\right)^4
\fr{a+a_{eq}}{a^{(0)}}.
\label{vofphi}
\ee
In this expression, $H^{(0)}$ and $1-\Omega^{(0)}_\phi = \Omega^{(0)}_m$ are the 
Hubble expansion rate and the matter density today. 
This equation can be used further to obtain $\phi$, {\em i.e.} by taking the logarithm of
both sides of (\ref{vofphi}) for a simple exponential potential. The derivative of 
(\ref{vofphi}) with respect to $x$ gives us another useful equation, 
\be 
\frac{d \ln(1 - \omega_{\phi})}{d x} = 
3( 1 +
\omega_{\phi} ) + \frac{\partial \ln V}{\partial \phi} \frac{d
\phi}{dx}.
\label{domega} 
\ee

When the change in  $B_m(\phi)$ is not small and cannot be neglected, 
all cosmological equations become considerably more complicated. 
The scaling of the matter energy density differs from usual $a^{-3}$ behavior 
because of the changing mass, due to $B_m(\phi)$, while the scaling of radiation 
energy density remains unchanged,
\begin{eqnarray}
\rho_{m}(x) &=& \rho_{m}^{(0)} a^{-3} \frac{B_{m}\Bigl(\phi(x)
\Bigr)}{B_{m} \Bigl(\phi(0) \Bigr)},~~~{\rm and}
~~~ \rho_{r}(x) = \rho_{r}^{(0)} a^{-4} \label{rhor2} 
\label{rhoi2} 
\\
\rho_{r} &=& \frac{a_{eq}}{a} \frac{B_{m}\Bigl(\phi(x_{eq})
\Bigr)}{B_{m} \Bigl(\phi(x) \Bigr)} \rho_{m} 
\end{eqnarray}
With the use of these relations, we generalize Eqs. (\ref{dOmega}) and (\ref{domega}),
\be 
\frac{d \ln(1 -\Omega_{\phi})}{d x} =  \Omega_{\phi}
\Biggl[3 \omega_{\phi} - \Bigl(\frac{a_{eq}^{(c)}}{a+
a_{eq}^{(c)}} \Bigr) \Biggr] + \Bigl(\frac{a}{a+ a_{eq}^{(c)}}
\Bigr) \frac{\partial \ln B_{m}}{\partial \phi} \frac{d \phi}{dx}
\label{dOmega1} 
\ee
\be
\frac{d \ln(1 - \omega_{\phi})}{d x} =  3( 1 +
\omega_{\phi} ) + \frac{\partial \ln V}{\partial \phi} \frac{d
\phi}{dx} + \frac{(1 - \Omega_{\phi})}{\Omega_{\phi}}
\Bigl(\frac{a}{a+ a_{eq}^{(c)}} \Bigr) \frac{\partial \ln
B_{m}}{\partial \phi} \frac{d \phi}{dx} 
\label{domega1} 
\ee
where we have introduced an auxiliary function,
$a_{eq}^{(c)}(\phi) = a_{eq} B_{m}\Bigl(\phi(x_{eq})
\Bigr)B^{-1}_{m} \Bigl(\phi(x) \Bigr)$.

\section{Cosmological evolution of the scalar field}
\setcounter{equation}{0}

\subsection{Relation between $V(\phi)$ and $B_i(\phi)$}

Our main assumption in this work is  that all functions $B_i(\phi)$ and $V(\phi)$ admit
a common extremum, which is an obvious generalization of Damour-Nordtvedt and 
Damour-Polyakov constructions \cite{DN,DP}. Near this extremum, all functions admit an expansion
\be
B_i(\phi) = 1 + \fr{1}{2}\xi_i \phi^2+...;\;\;\;\;\;\;\;
V(\phi) = V_0( 1 +  \fr{1}{2}\lambda \phi^2+...),
\label{dp}
\ee
where $\xi_i$ and $\lambda$ are dimensionless parameters, while $V_0$ is of the
order of the dark energy density today.

In order to be able to start our analysis, we must specify
the potential $V(\phi)$ and functions $B_i(\phi)$. 
Without any serious guidance from the underlying theory, their choice is 
completely {\em ad hoc}, and the only constraints are those coming 
from observations. A large number 
of quintessence potentials $V(\phi)$ has been analyzed in the past,
but the parameter space of functions $B_m(\phi)$ and $B_F(\phi)$ 
remains relatively unexplored. To this end, we propose an ansatz
that would relate $B_{i}(\phi)$ and $V(\phi)$. Since we are going to consider 
only positive-definite functions, we choose a two-parameter relation, 
\be 
B_{i}(\phi) = \Biggl(\frac{b_i+V(\phi)/V_0}{1+b_i}\Biggr)^{n_i}, ~~~ {\rm where}~~ b_i+1>0
\label{BF2} 
\ee 
This relation is sufficiently generic, and varying the two dimensionless 
parameters $n_i$ and $b_i$ for each function $B_i(\phi)$ allows us to cover a 
wide range of possibilities. The expansion of (\ref{BF2}) near the 
extremum allows us to relate $\xi_i$ and $\lambda$,
\be
\xi_i = \fr{\lambda n_i}{1+b_i}.
\ee
If the quintessence field changes by $\Delta \phi \sim O(1)$ between 
redshifts $z\sim 1$ and $z=0$, the constraints on the variation of masses 
and couplings would normally require small $\xi_i$ which may be due to
small $n_i$ and/or a large value for $b_i$. 
In a typical tracker regime, $V(\phi)$ changes by many orders of magnitude, and
at early cosmological epochs, the value of $V(\phi)$ can be much larger than 
the normalization scale $b_i$. In this case, $B_i(\phi)\sim (V(\phi))^n$. 

In the next two sections, we analyze two types of potentials, which in the 
asymptotic regime of tracking behavior, scale as $\exp(\phi^2)$ and $\exp(\phi)$.
We first assume that the couplings of $\phi$ to all matter fields are 
rather small and have no backreaction on the cosmological evolution of $\phi$. 
We then allow for a substantial coupling $B_m(\phi)$ to dark matter and show the
impact of that on the dynamics of the quintessence field. By comparing the evolution 
of $B_F(\phi)$ for electromagnetic field to observational constraints, we can determine 
the allowed range of parameters for $n_F$ and $b_F$. 


\subsection{Quintessence with $\exp(\lambda\phi^2/2)$ and $\cosh(\lambda\phi)$ type potentials}

We next consider two sample potentials which can combine the properties of 
a tracking solution with a minimum that provides the dark energy at 
late times:
\begin{eqnarray}
\label{Vphi}
{\rm case ~~A}: ~~~~~~~V(\phi) &=& V_{0} \exp \Bigl( \frac{\lambda\phi ^2 }{2} \Bigr)\\
{\rm case ~~B}: ~~~~~~~V(\phi) &=& V_{0} \cosh(\lambda\phi)
\label{Vphi1}
\end{eqnarray}
Both potentials have the same expansion near the minimum (\ref{dp}) and are sufficiently 
easy to analyze. For now, we set $B_F = B_m = 1$. 
Closely related potentials have been 
expansively analyzed in the past by several groups 
\cite{Brax,Sahni}.

{\em Case A.} 

While the evolution equation for $\Omega_\phi$ is unchanged from
Eq. (\ref{dOmega}), the evolution equation for $\omega_\phi$ can be
rewritten as
\be 
\frac{d \ln (1 - \omega_{\phi})}{d x} =
 3 (1 + \omega_{\phi}) \pm \lambda \phi
\sqrt{3 \Omega_{\phi} (1 + \omega_{\phi})},
\label{domega3} 
\ee 
where the choice of plus(minus) depends on whether
the evolution of $\phi$ starts from negative(positive) values. 
The dependence of $\phi$ can be obtained directly from (\ref{vofphi}), 
\be 
\fr\lambda2\phi^2 = - 4x + \ln(a+a_{eq})+
\ln\left(\fr{\Omega_\phi(1-\omega_\phi)}{1-\Omega_\phi}\right)+
\ln \left[ \frac{3(H^{(0)}\bar M)^2(1-\Omega_\phi^{(0)})}{2V_0} \right] 
\label{phi3} 
\ee 
All terms in  Eq.
(\ref{phi3}) other than the $(-4 x)$-term do not change
significantly over the history of Universe. This is similar to
the evolution of a scalar field in the tracker regime for an exponential
potential, where $\ln V(\phi) \sim {\rm const} - 
3 (1 + \omega_{i}) x$ \cite{Ng}.

An advantage of a tracker-type potential is that 
the cosmological evolution of the scalar field $\phi$
during the observationally relevant epoch, {\em i.e.} $z\la 10^{10}$
is insensitive to the choice of initial conditions for $\phi$
that can be specified deep inside the radiation-dominated epoch.
In order to exhibit tracking behavior, the
potential must satisfy two conditions \cite{Steinhardt}: 
\be
\Gamma \equiv \fr{V''V}{(V')^2} \ge 1 ~~{\rm and}~~ \left|\fr{d\ln(\Gamma - 1)}{dx}
\right| \ll 1. 
\label{tc} 
\ee 
The simple exponential potential provides a tracker solution
with $\Gamma = 1$.

For the potential (\ref{Vphi}), we find 
\be 
\Gamma = 1 +
\frac{1}{\lambda \phi^2},
\label{Gamma} 
\ee 
which is larger than 1 for any positive $\lambda$. 
For the second condition we obtain
\be 
\left|\frac{d \ln (\Gamma -1)}{dx}\right| = 
\frac{2\sqrt{3\Omega_{\phi} ( 1 + \omega_{\phi})}}{|\phi|},
\label{Gammaslow} 
\ee 
which is much smaller than one as long as $\phi$ is far from the 
minimum, {i.e.} at all cosmological epochs except for late times.

\begin{center}
\begin{figure}
\vspace{1cm}
\centerline{
\psfig{file=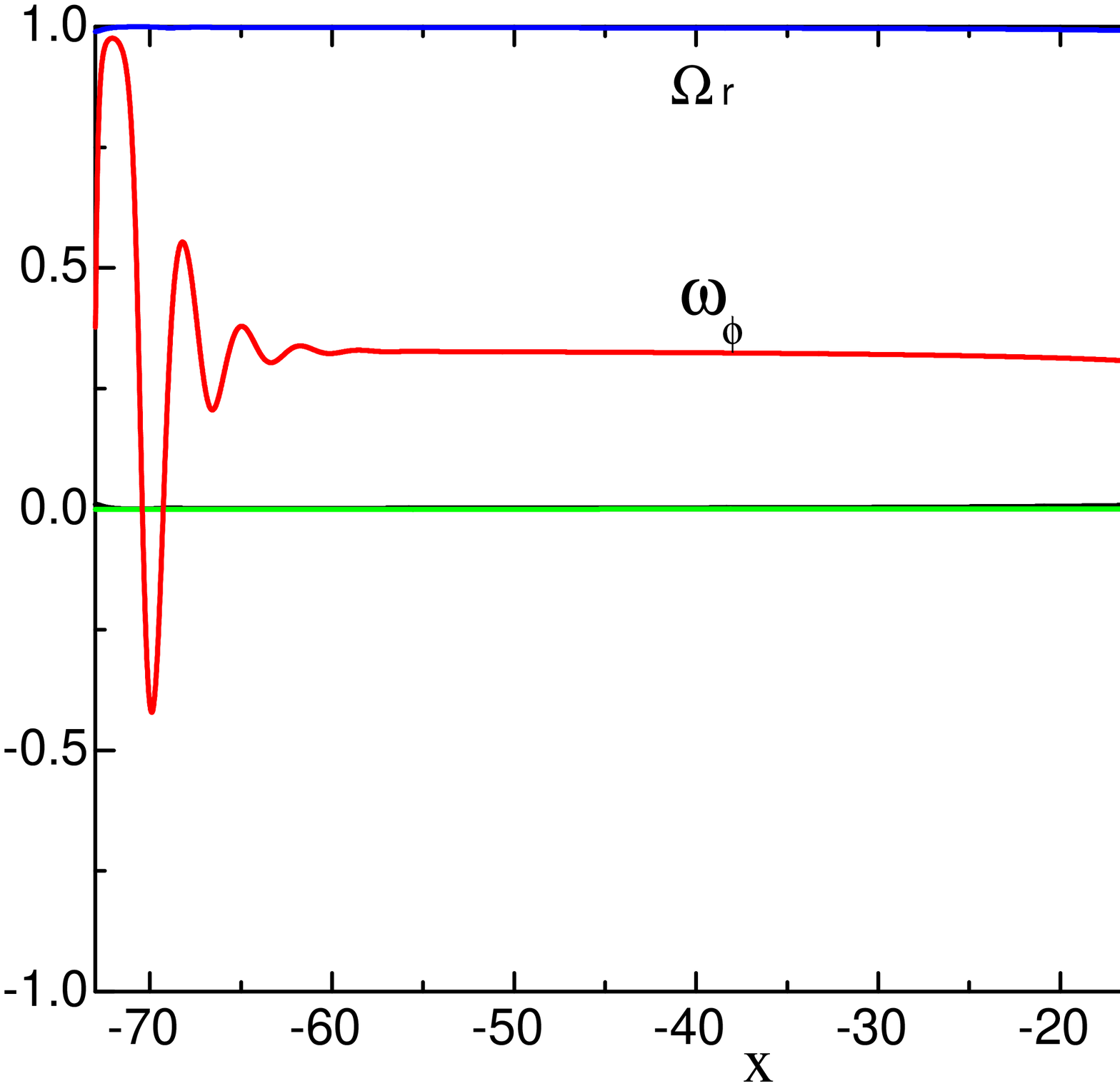, width=8cm}\psfig{file=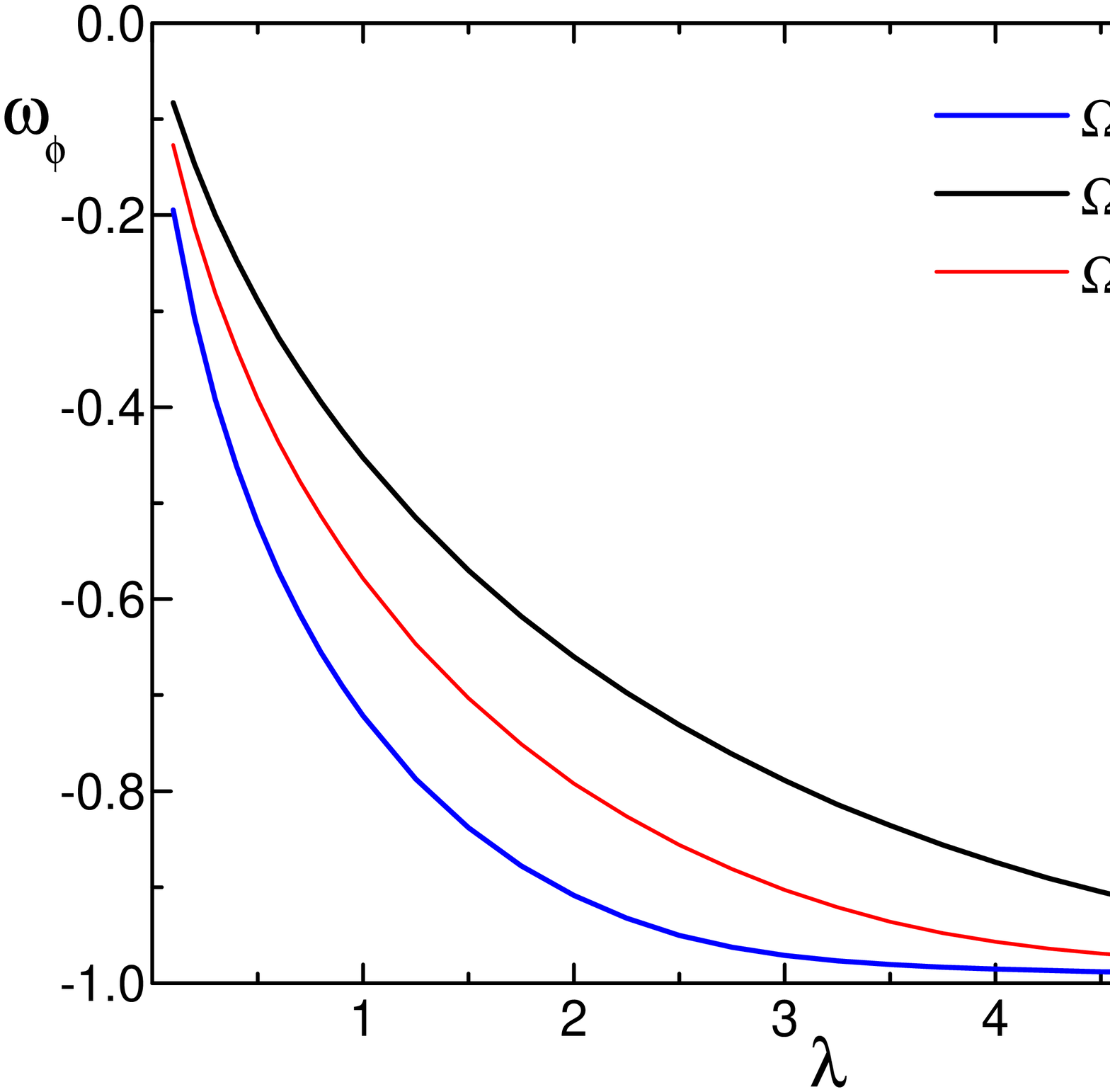, width=8cm} }
\vspace{-1cm}
\caption{ The $\exp(\lambda\phi^2/2)$ potential.
a) The cosmological evolution of the equation of state parameter,
$\omega_{\phi}$, and the energy density parameters, $\Omega_{i}$,
of each component for $\lambda = 5$. The evolution of the parameters is similar
for other choices of $\lambda$. 
b) The dark energy equation of state parameter $\omega_\phi$ as a 
function of $\lambda$ for an acceptable range of the 
dark energy density ($0.6 \leq \Omega_{\phi}^{(0)} \leq 0.85$).}
 \label{fig1}
\end{figure}
\end{center}

To be compatible with observational data, quintessence models 
must satisfy the following conditions:  1) The
energy density of quintessence must be subdominant during Big Bang nucleosynthesis
(BBN) \cite{FJ},
$\Omega_{\phi}^{(BBN)}(x
\sim -23) < 0.2$  at $T \sim 1$ MeV; 2)
the energy density at present must be compatible 
with the preferred range for dark energy, $0.60 \leq
\Omega_{\phi}^{(0)}(x = 0) \leq 0.85$ \cite{wmap}.
The evolution of energy density components, $\Omega_r, \Omega_m$, and
$\Omega _\phi$ is shown in Fig. \ref{fig1}a along with the dark energy
equation of state. 
In the very early universe, the EOS parameter has an oscillatory behavior as
does the density parameter $\Omega_{\phi}$ which quickly converges 
to an attractor solution washing away any memory of initial conditions. 
As one can see, the constraint of
$\Omega_{\phi} < 0.2$ from nucleosynthesis is
easily satisfied in this model. We also note that during the
radiation dominated era,  $\omega_{\phi} \approx \omega_{r} = 1/3$
as in other tracker solutions. The transition to a
matter dominated Universe occurs around $x = -8.6$ ($z \sim 5400$). 
During the matter dominated stage, $\omega_\phi$ begins to drop sharply
towards its final value of $\omega_\phi = -1$.  At $x = -0.5$ ($z \simeq 0.6$), the Universe 
begins to be dominated by the quintessence field.
In Fig. \ref{fig1}b, we show the present value of 
$\omega_{\phi}$ as a function of the coupling,
$\lambda$, for three different values of the present value of $\Omega_\phi$. 
For large values of $\lambda$, the
potential is relatively steep and the field quickly evolves
towards  its minimum where $V\simeq V_0$, which 
must be tuned close to the present cosmological constant value. 
In this case, $\omega_\phi \to -1$. In contrast, for small values of
$\lambda$, the potential is relatively shallow, and $\phi$ is
still evolving.  As one can see, applying the constraint
\cite{PTW}, $\omega_{\phi}^{(0)} < -0.6$ gives the limit
$\lambda \ga 1.5$ (for the case of $\Omega_\phi = 0.73$). 
When the present value of $\phi$ is not at the minimum ($\phi = 0$),
the constant, $V_0$ must be adjusted to obtain the desired value of
$\Omega_\phi$ today.

{\em Case B}.

\begin{center}
\begin{figure}
\vspace{1.5cm}
\centerline{\psfig{file=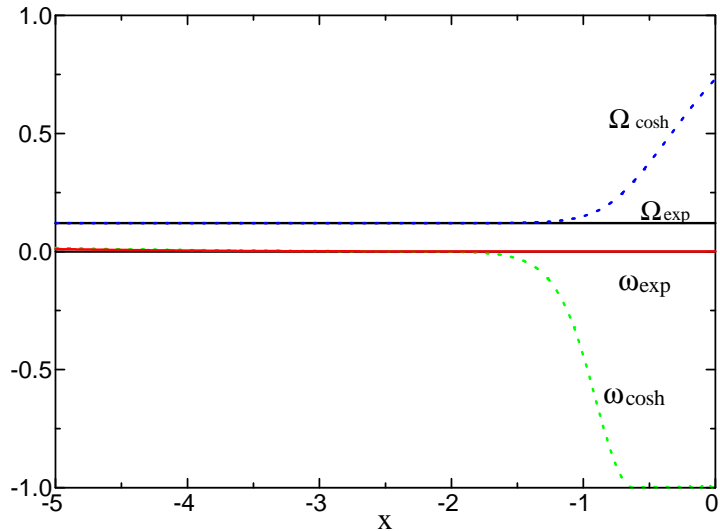, height=7cm}}
\vspace{-1.5cm}
\caption{ The evolution of $\Omega_\phi$ and $\omega_\phi$ for
the simple exponential 
potential (solid) and the cosh potential (dotted). The main difference
between two potentials appears at late times. 
We use $\lambda =5$ for both potentials.
} \label{fig4}
\end{figure}
\end{center}

As one might expect, the potential (\ref{Vphi1}) gives a very
similar result as one would obtain for a simple exponential potential 
($V(\phi) \propto \exp(-\lambda \phi)$) at early times.
Therefore, for $\lambda\phi \ga 1$, {\em i.e.} far from the minimum, 
the scalar field is simply 
\be
\lambda\phi \simeq - 4x + \ln(a+a_{eq})+
\ln\left(\fr{\Omega_\phi(1-\omega_\phi)}{1-\Omega_\phi}\right)+
\ln \left[ \frac{3(H^{(0)}\bar M)^2(1-\Omega_\phi^{(0)})}{2V_0} \right].
\label{phi4}
\ee
For this potential, the
tracking parameter $\Gamma$ is,
\be 
\Gamma = 1 + \frac{1}{\sinh^2(\lambda \phi)} .
\label{Gamma2} 
\ee 
Clearly, at late times, as $\phi$ approaches $0$, 
the tracking condition will be violated which opens up the possibility
that the Universe will be dominated by the scalar field, in contrast to the
simple exponential potential. 
A comparison of the late time behavior between the exponential and
cosh type potentials is shown in Fig.\ref{fig4}.

\begin{center}
\begin{figure}
\vspace{1cm}
\centerline{
\psfig{file=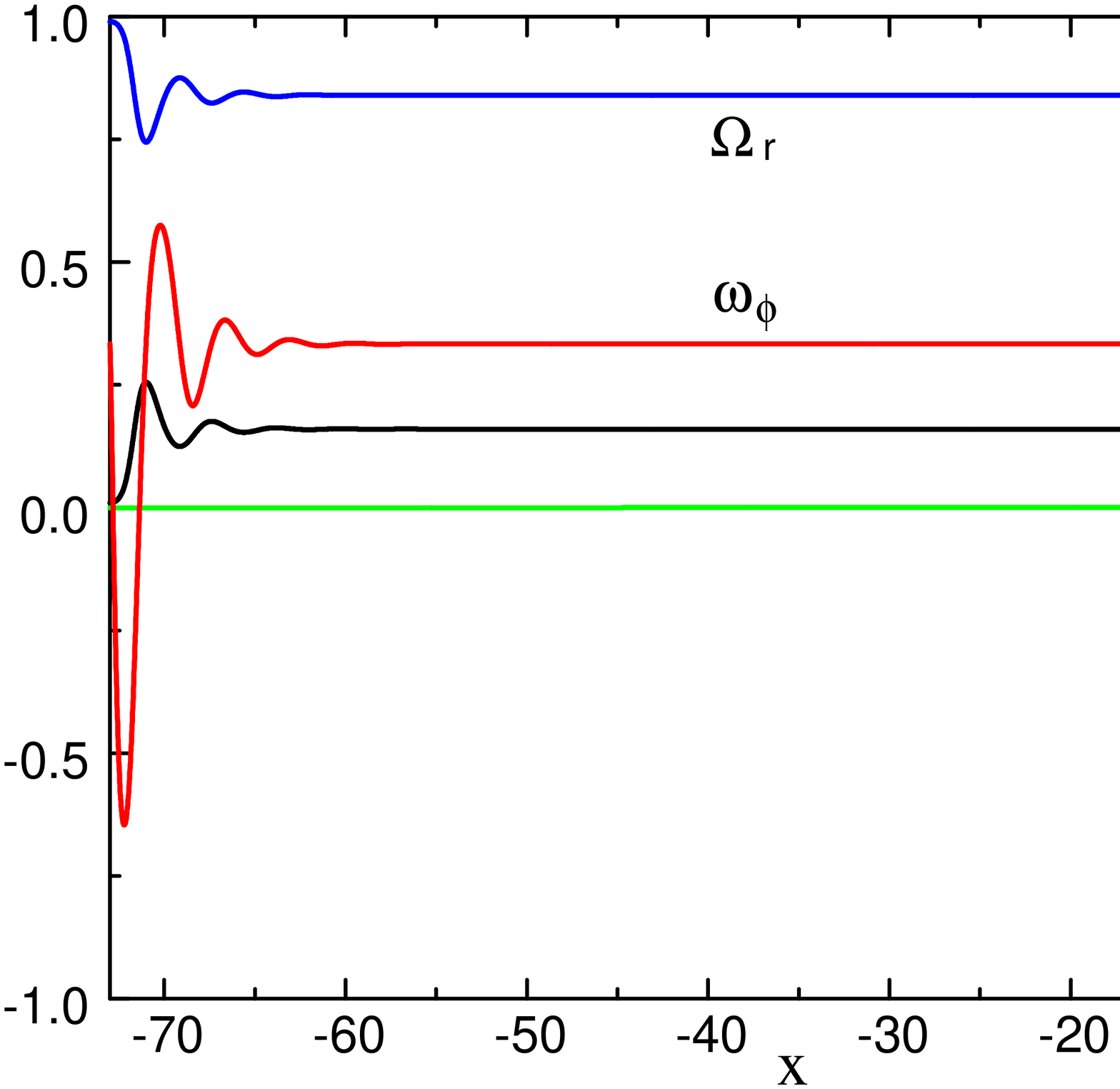, width=8cm}\psfig{file=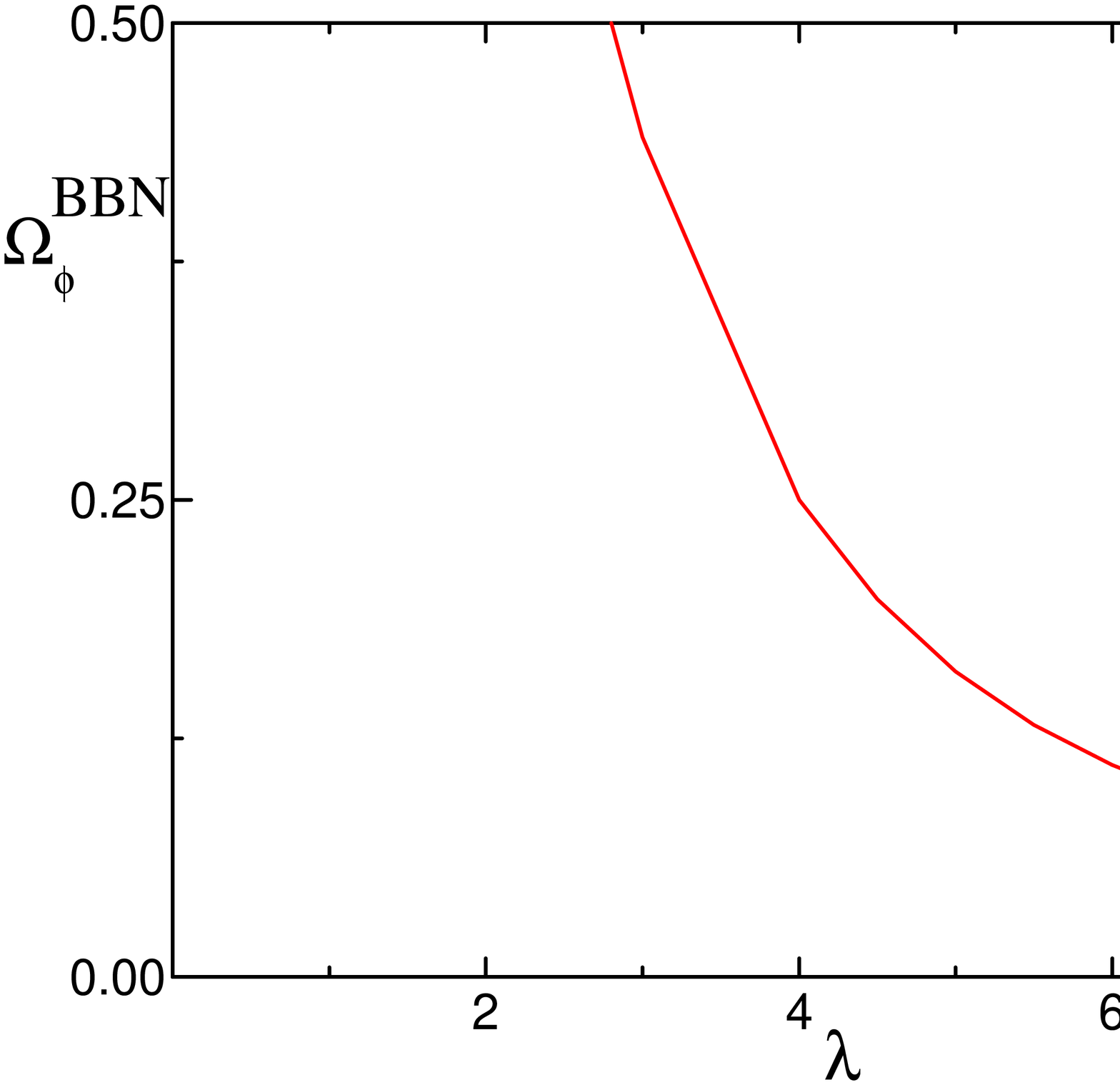, width=8cm} }
\vspace{-1cm}
\caption{The $\cosh (\lambda \phi)$ potential.
a) The cosmological evolution of the equation of state parameter,
$\omega_{\phi}$, and the energy density parameters, $\Omega_{i}$,
of each component for $\lambda = 5$. The evolution of the parameters is similar
for other choices of $\lambda$. 
b) The quintessence density parameter  $\Omega_\phi$ evaluated at the
time of BBN as a 
function of $\lambda$. }
 \label{cosh}
\end{figure}
\end{center}

In Fig. \ref{cosh}a, we show the full evolution of 
$\Omega_r, \Omega_m$, 
$\Omega _\phi$, and $\omega_\phi$.
In this case, matter domination occurs at $x = -8.2 $ ($z \simeq 3500)$, 
and quintessence domination occurs at
$x = -0.3$ ($z \simeq 0.35$).
The cosh-type potential turns out 
to be more restricted by observational requirements than the potential (\ref{Vphi}).   
In Fig. \ref{cosh}b, we show the value of $\Omega_\phi(x = -23)$ corresponding
to its BBN value as a function of the coupling $\lambda$.  
We see that the constraint  $\Omega_\phi < 0.2$ implies
$4\leq \lambda$.

\section{Time variation of the fine structure constant 
}
\setcounter{equation}{0}

To study the cosmological evolution of the fine structure constant we use the following
relation between $B_F$ and $\alpha$,
\be
\fr{\Delta \alpha(z)}{\alpha} \equiv  \fr{\alpha(z)-\alpha(0)}{\alpha(0)}
= \fr{B_F(\phi(0))}{B_F(\phi(z))}-1.
\label{BF}
\ee
Note that the present value of the field $\phi(0)$ is close to zero. 

We assume that the backreaction of $B_i(\phi)$ on the cosmological evolution of 
$\phi$ itself is small, and therefore can use the results of the previous section. 
All physical processes sensitive to changes in $\alpha$ can be separated into two 
broad categories:
those that correspond to relatively low redshifts, $z \sim 1$ and smaller, and 
high-redshift phenomena such as BBN and the cosmic microwave background anisotropies.

\subsection{Late time evolution}

For the late time evolution of $\alpha$ we can use the expansion 
of $B_F(\phi)$ given in (\ref{dp}). In this case, the expression for 
$\Delta(\alpha)$ takes the following simple form,
\be
\fr{\Delta \alpha(z)}{\alpha} = \fr{\xi}{2}\left(\phi^2(0)-\phi^2(z)\right),
\label{alphaexp}
\ee
where we dropped the subscript $F$ in $\xi_F$ to be concise. 
For the exponential of $\phi^2$ potential (\ref{Vphi}), this expression can
be found in a more explicit form with the use of (\ref{phi3}), 
\be
\fr{\Delta \alpha(x)}{\alpha} = \fr{\xi}{\lambda}\left[
- 3x + \ln\left(\fr{\Omega_\phi(1-\Omega^{(0)}_\phi)(1-\omega_\phi)}
{\Omega^{(0)}_\phi(1-\Omega_\phi)(1-\omega_\phi^{(0)})}\right)
\right].
\label{dalpha1}
\ee

Using the result of the previous section, we can predict the evolution 
of $\alpha$ over redshift in terms of two parameters, $\xi$ and $\lambda$. 
We choose two characteristic values of $\xi$, based on two QSO results.
To be consistent with the non-zero result for $\Delta \alpha$ 
by Murphy et al. \cite{Webb}, we choose $\xi $ in such a way that 
$\Delta \alpha/\alpha = -5.4 \times 10^{-6}$ at a redshift of 3. 
Another option that we explore is $|\Delta \alpha/\alpha| \leq 6\times 10^{-7}$
at $z=1.5$, which is motivated by the experimental accuracy of Chand et al. \cite{Petitjean}.
For definiteness, in the second case we choose $\Delta \alpha/\alpha =-7\times 10^{-6}$. 

\begin{center}
\begin{figure}
\vspace{1cm}
\epsfxsize=5.5cm \centerline{\psfig{file=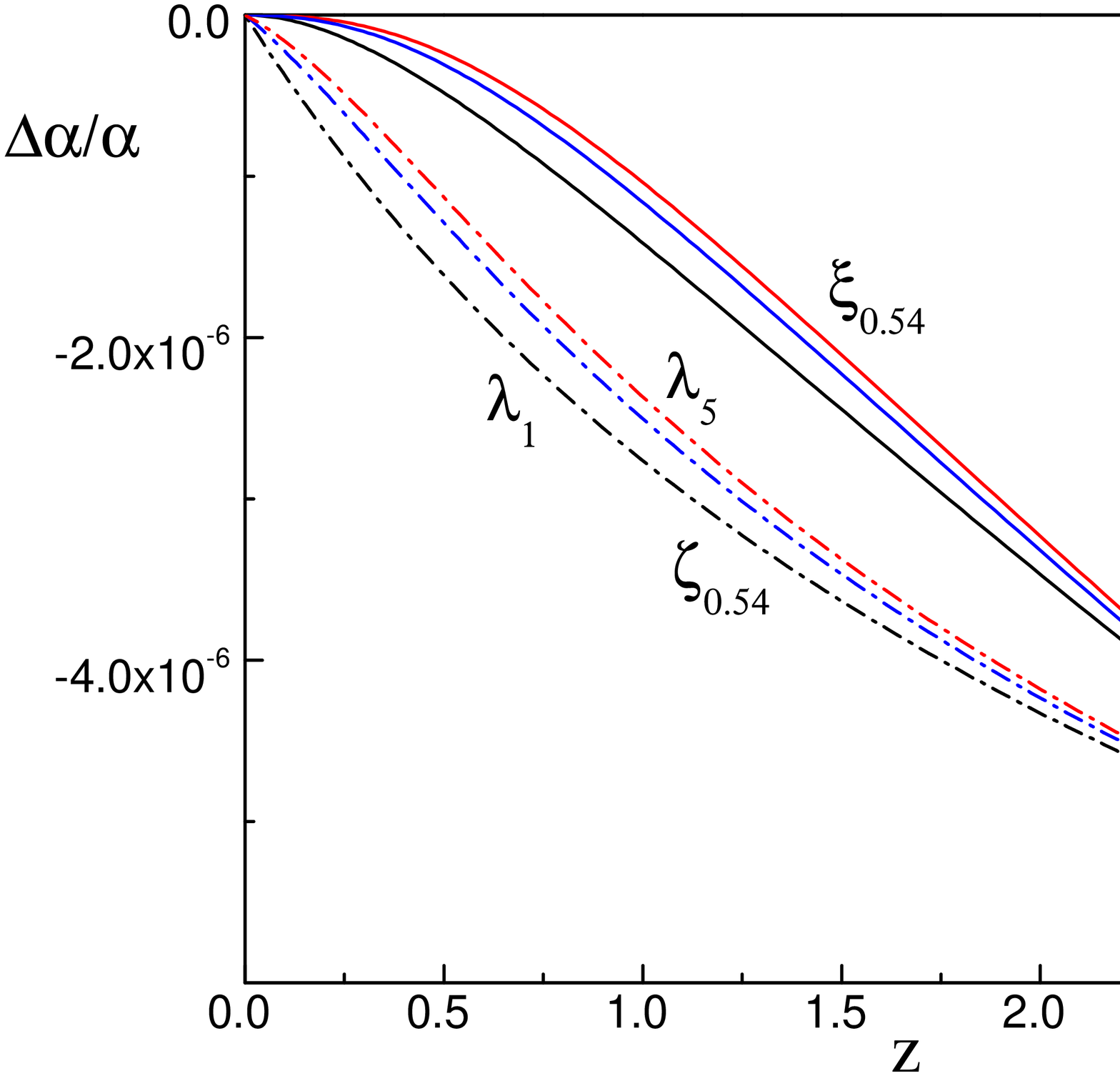, height=6cm}\psfig{file=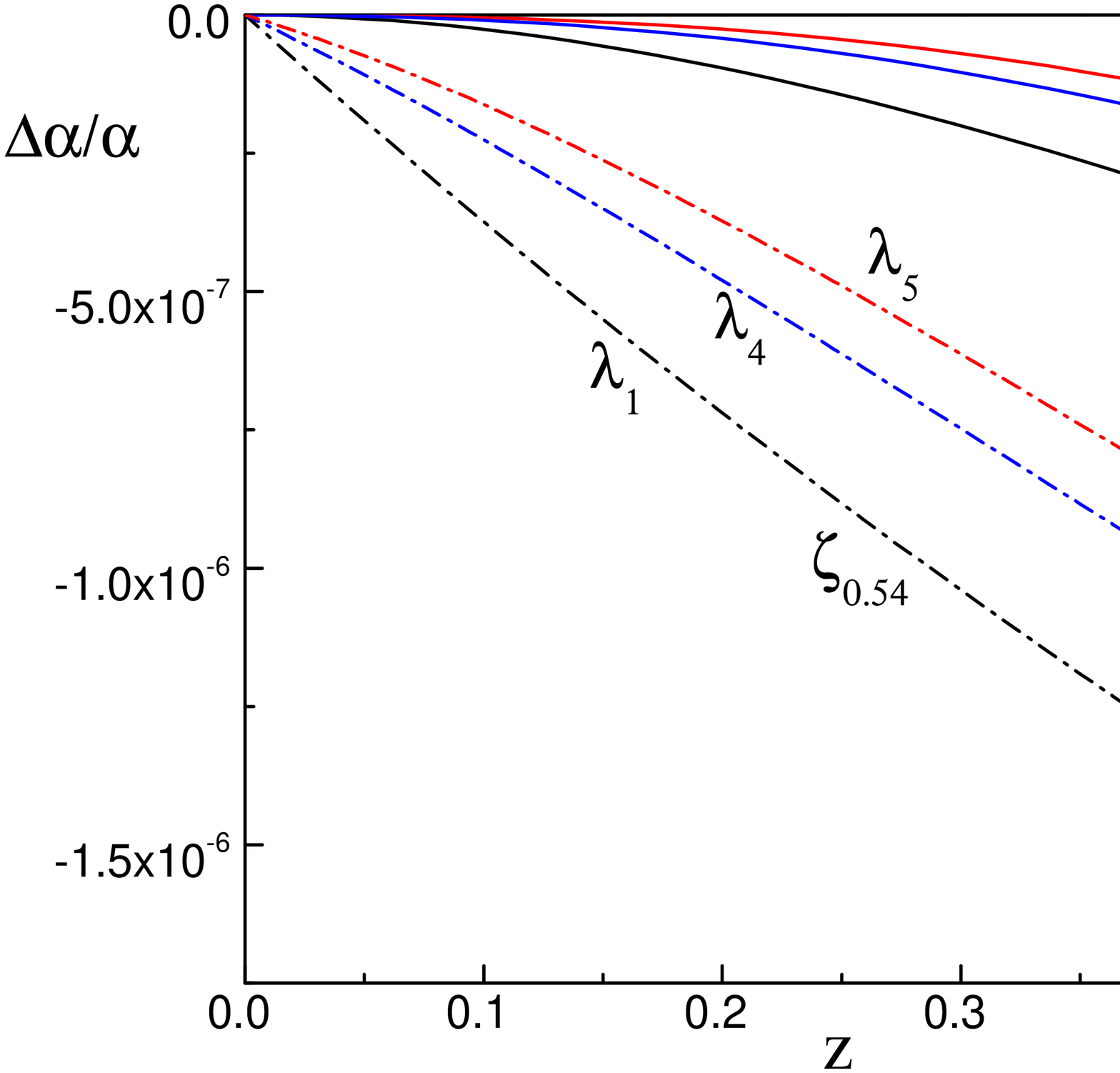, height=6cm}}
\epsfxsize=5.5cm \centerline{\psfig{file=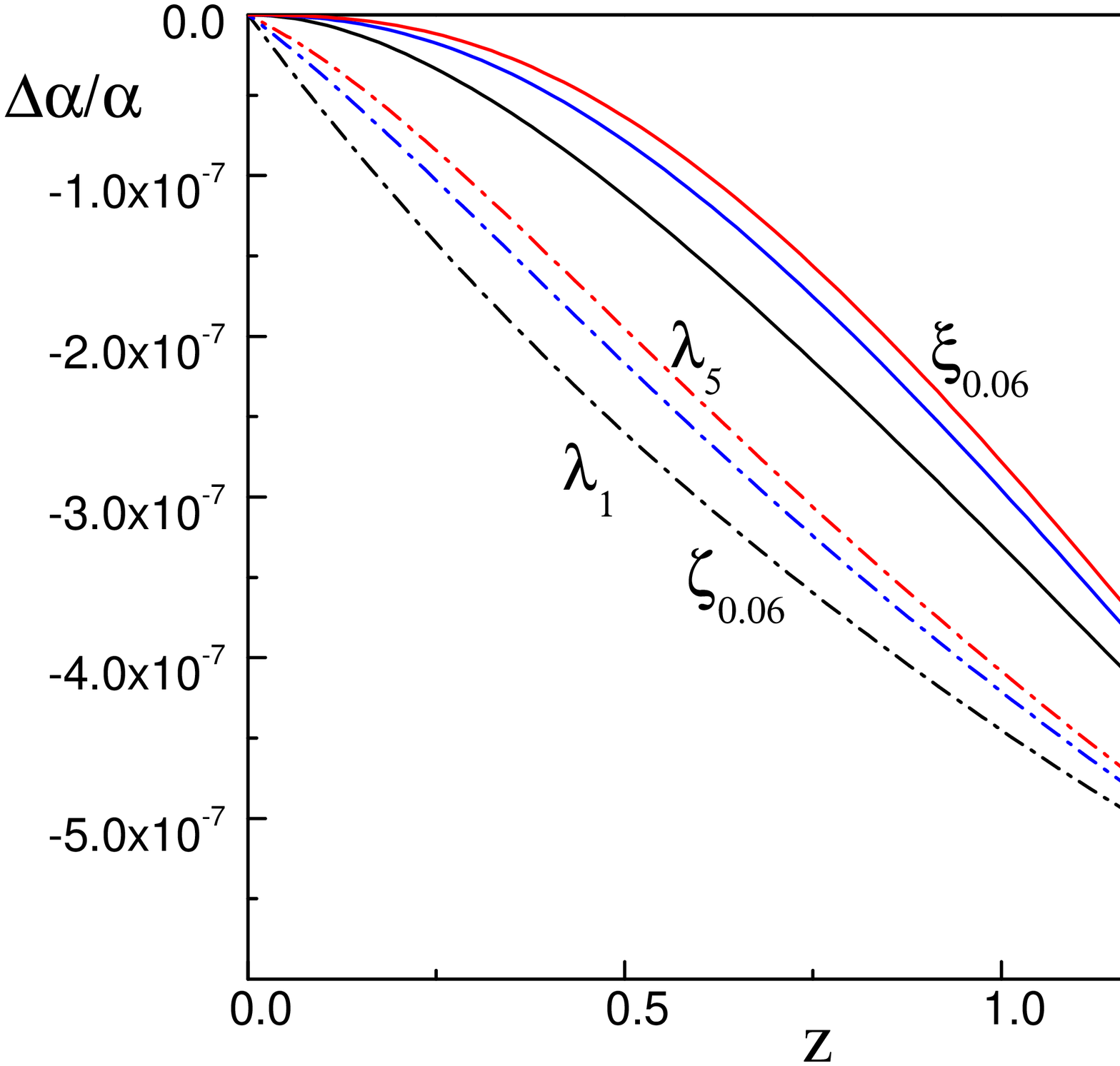, height=6cm}\psfig{file=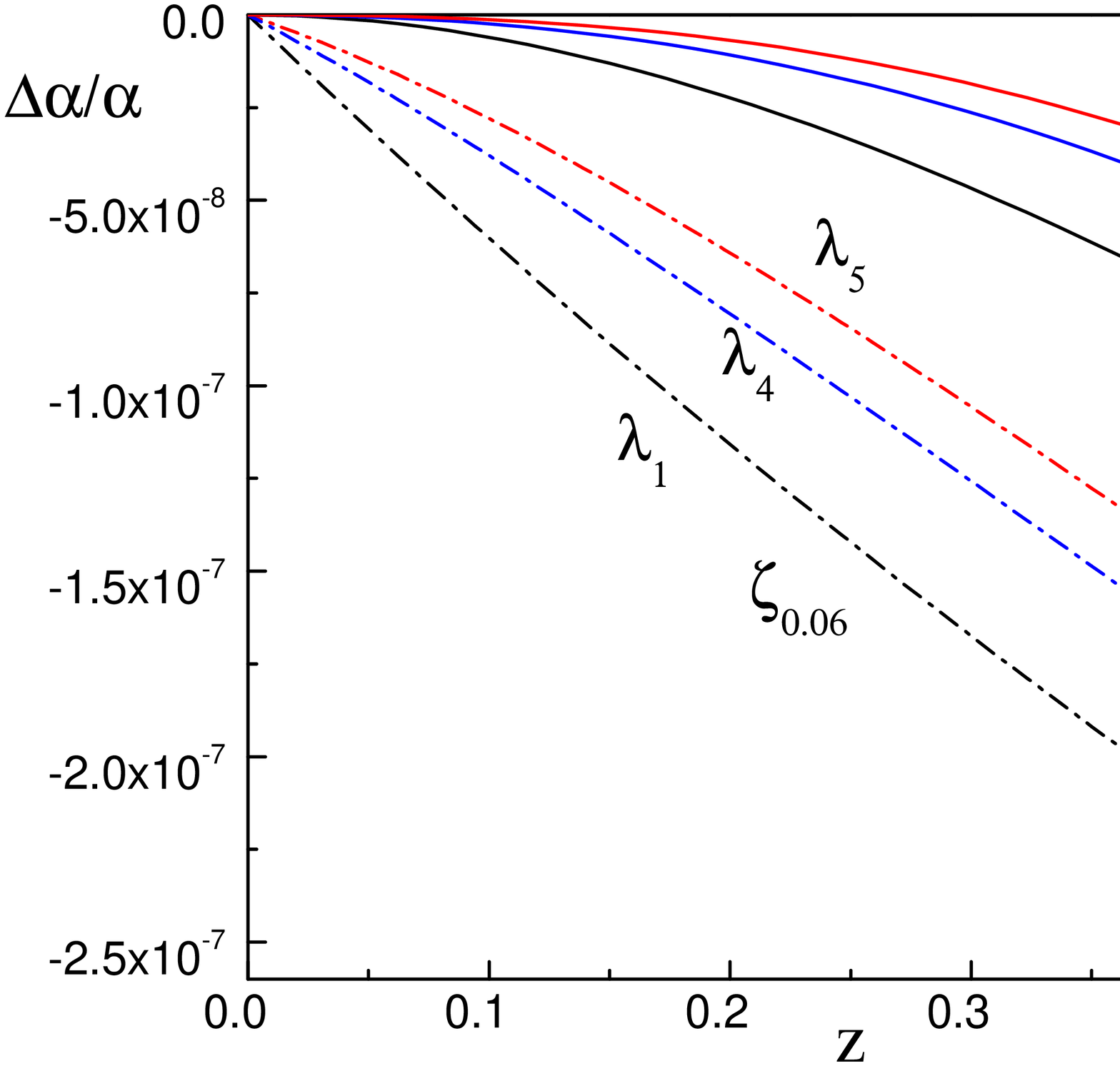, height=6cm}}
\vspace{-1cm}
\caption{ a. The evolution of $\Delta\alpha/\alpha$ over the redshift range
$0\leq z\leq 3$ driven by the potential (\ref{Vphi}). Panels a) and b) use the common 
normalization $\Delta \alpha/\alpha = -0.54 \times 10^{-5}$ at $z=3$. 
Figures c) and d) use the common 
normalization $\Delta \alpha/\alpha = -0.06 \times 10^{-5}$ at $z=1.5$. 
The solid lines correspond to the choice $\zeta = 0$ and $\xi \ne 0$, whereas 
the dashed-dotted lines allow $\zeta \ne 0$.  
} \label{alphaV1}
\end{figure}
\end{center}

The results for the choice $\Delta \alpha/\alpha = -5.4 \times 10^{-6}$ are 
plotted in Fig. \ref{alphaV1}. 
The family of solid 
curves corresponds to different choices of the coupling $\lambda = 1, 4$, and $5$ in (\ref{Vphi}). 
It is easy to see that 
the evolution of $\alpha$ is approximately 
linear over a large range of redshifts for  $z>1$. This follows directly 
from (\ref{dalpha1}) where only the $-3x$ term is changing at these redshifts. 
In panel b) of Fig. \ref{alphaV1} we show
a blow-up of panel a) for  $z\leq 0.5$. This will facilitate the comparison
between this model and the experimental constraints at low $z$.
At late times (at low $z$), the evolution
of $\alpha$ slows down considerably. For comparison, we present similar results
for the case of the linear coupling $\zeta$ between $\phi$ and $F_{\mu\nu}F^{\mu\nu}$,  for
the same choice of initial conditions. The corresponding curves are plotted in Fig. \ref{alphaV1}
by dashed-dotted lines. 
The main qualitative difference between these two choices, $\xi$ or $\zeta$ is at late
times. When $\zeta = 0$ and $\xi \ne 0$, the fine structure constant evolves very slowly,
whereas for $\zeta \ne 0$
the evolution of $\alpha$ remains approximately linear. 

\begin{center}
\begin{figure}
\vspace{1cm}
\epsfxsize=5.5cm \centerline{\psfig{file=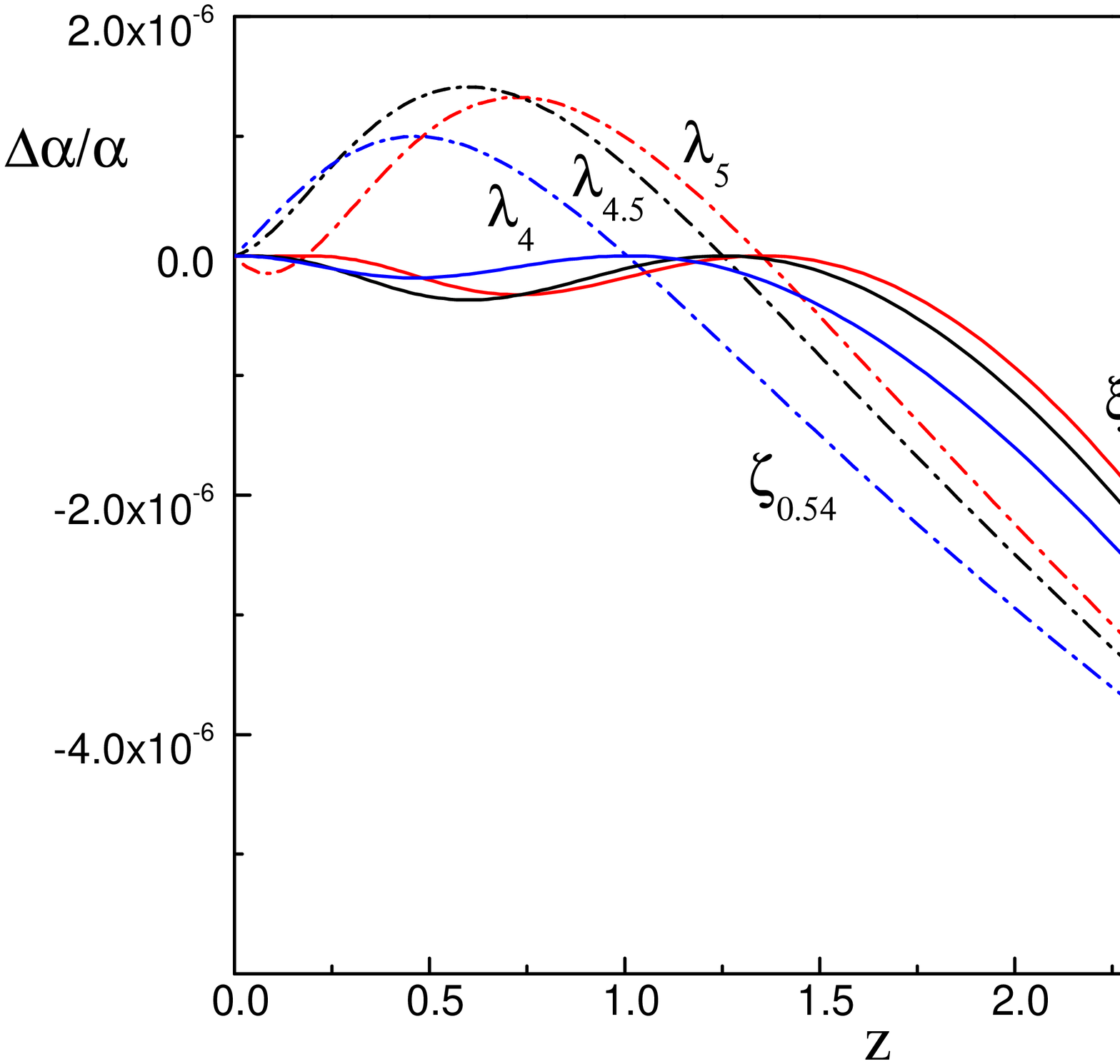, height=6cm}\psfig{file=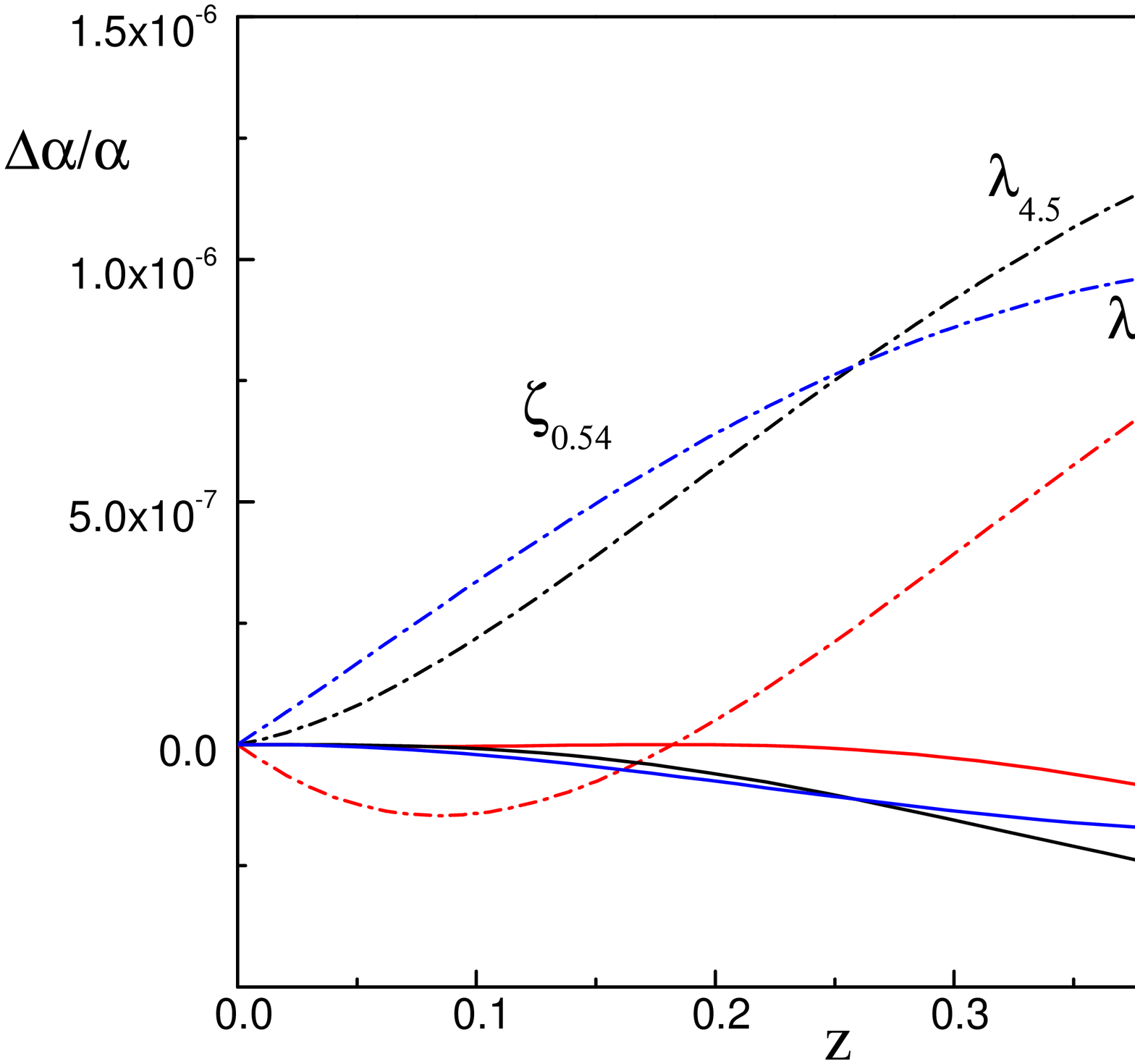, height=6cm}}
\epsfxsize=5.5cm \centerline{\psfig{file=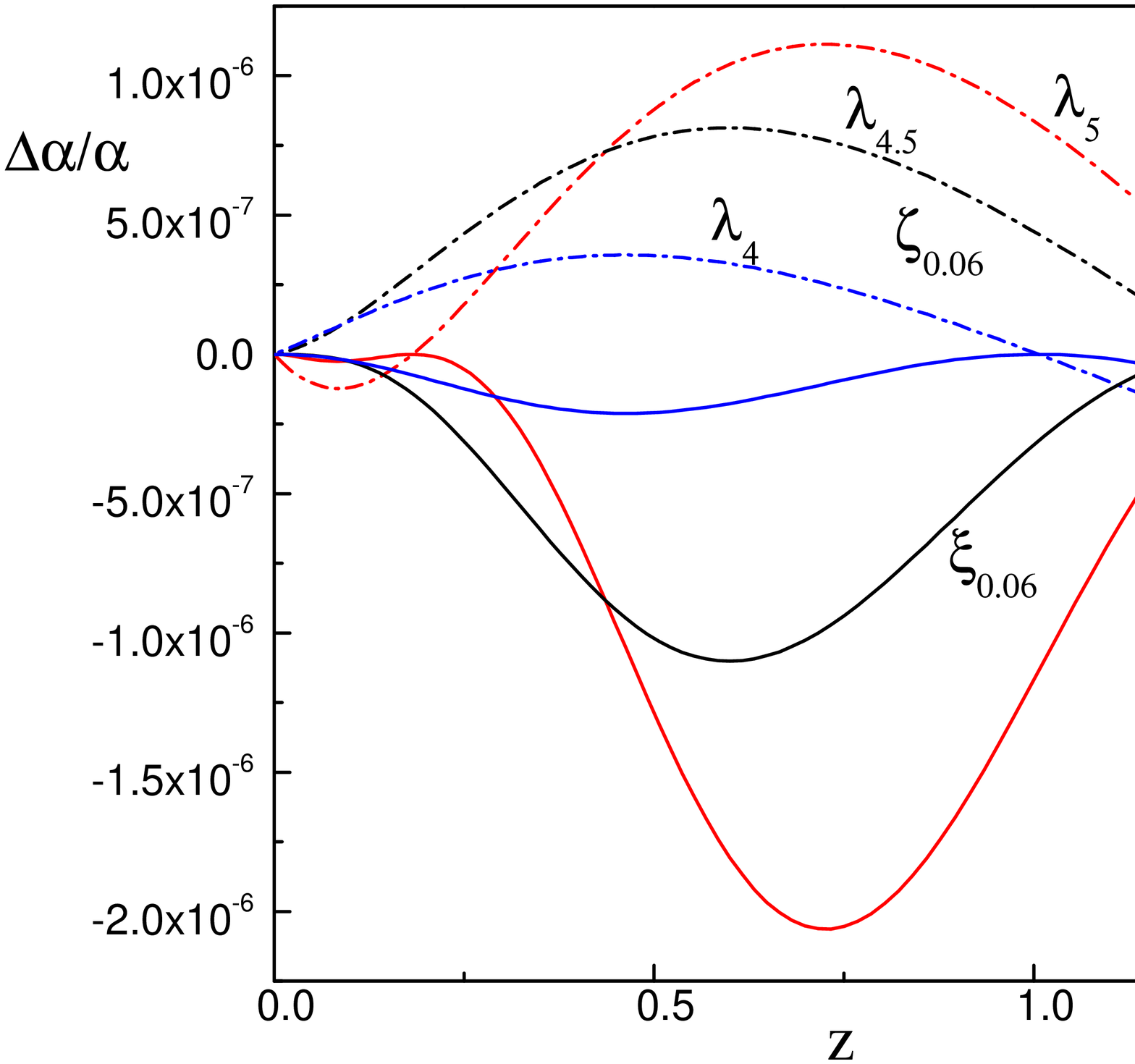, height=6cm}\psfig{file=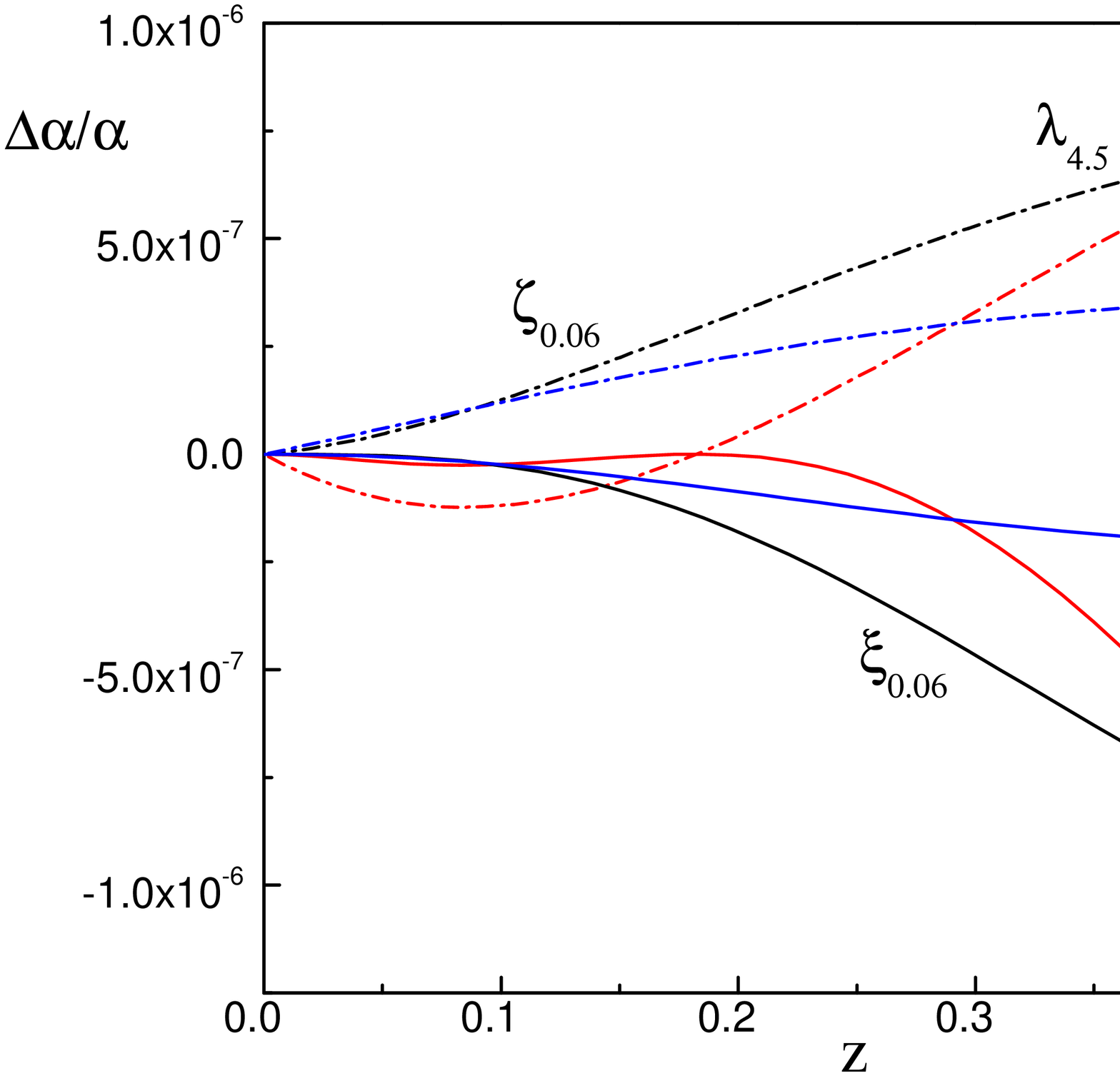, height=6cm}}
\vspace{-1cm}
\caption{ As in Fig. \protect\ref{alphaV1} for the potential \protect\ref{Vphi1}.
Here $\lambda = 4, 4.5$, and 5.
} \label{alphaV2}
\end{figure}
\end{center}

In Fig. \ref{alphaV2}, we show the analogous results for the evolution of $\alpha$ for the
potential (\ref{Vphi1}).  Once again, we show results for three choices of $\lambda$; in this
case, we have used $\lambda = 4, 4.5$, and 5. Notice that in this case, $\alpha$ overshoots
its present day value.  This is due to the fact that
with these choices of $\lambda$, $\phi$ rolls past the minimum (at $\phi = 0$) and begins
an oscillatory behavior as it settles to the minimum of the potential.
For the quadratic coupling, variations in $\alpha$  do not change sign thus causing $\Delta
\alpha$ to reach the value 0, before becoming negative again.  Damped oscillations
of $\phi$ (and hence $\alpha$) will continue to settle towards the minimum 
for which $\Delta \alpha = 0$.

We now turn to the question of how the sensitivity to $\Delta\alpha/\alpha$ 
achieved in Refs. \cite{Webb,Petitjean} compares to other probes 
of $\Delta\alpha/\alpha$ at smaller redshifts. In particular, we 
calculate $\daa$ at $z=0.14$ relevant for the Oklo bound \cite{Oklo}, 
the time average $\overline{\Delta \alpha}/\alpha$ between the redshifts
0 and $z=0.45$ which is constrained  by meteoritic data \cite{OPQ,fuj}, and the rate of change
in $\alpha$ today, $\dot \alpha/\alpha$ \cite{fischer}. 
In addition to the observables directly related to the
fine structure constant, we also calculate the 
differential acceleration of two bodies towards a common attractor, 
limited by the precision tests of the universality of the gravitational 
force \cite{OP,EP}. To calculate $\overline{\Delta g}/g$ we use the calculation 
of Ref. \cite{OP}, and conservatively assume that the coupling of 
the scalar field to nuclei is mediated only by the electromagnetic interactions.
The calculations of Ref. \cite{OP} are done for the 
linear coupling $\zeta$, but the generalization to the case of 
$\zeta =0, \xi\neq 0$ is straightforward, via the substitution 
$\zeta \to \xi \phi(z=0)$ \cite{DP}.
The results of these calculations are compiled in Table \ref{tab:alpha}.

\begin{table}[htb]
\begin{center}
\caption{Results for $\Delta \alpha / \alpha$ for comparisons with the
Oklo, meteoritic, present-day, and equivalence principle bounds.
Values of $\Delta \alpha / \alpha$, $\xi$, and $\zeta$ have been scaled by
a factor of $10^6$, values of $\dot \alpha / \alpha$ and $\Delta g / g$ have
been scaled by a factor of $10^{17}$.
\label{tab:alpha}}
\vskip .3cm
\begin{tabular}{|c|c|c|c|c|c|c|c|c|}
\hline $\fr{V(\phi)}{V_0}$& $(\fr{\Delta\alpha}{\alpha})_{3} $ &
$(\fr{\Delta\alpha}{\alpha})_{1.5}$  & $\xi$ &$\zeta$&
$(\fr{\Delta\alpha}{\alpha})_{0.14}$ &
$(\fr{\overline{\Delta\alpha}}{\alpha})_{0.45}$ &
$\fr{\dot{\alpha}}{\alpha}$ (yr$^{-1}$) &
$\fr{\Delta g}{g}$ \\
\hline $\exp(\lambda \phi^2/2)$& $-5.4$ & $-3.4$ & $0$&
$-7.6$& $-0.24$& $-0.42$& $9.6$ & $-2.5$ \\
&$-5.4$& $-2.1$& $22$& $0$& $-0.011$&
$-0.05$& $0.048$ & $-0.074$ \\
&$-0.93$& $-0.6$& $0$& $-1.3$& $-0.041$&
$-0.073$& $1.7$ & $-0.076$ \\
&$-1.4$& $-0.6$& $5.8$& $0$& $-0.0028$&
$-0.013$& $0.013$ & $-0.0053$ \\
\hline $\cosh(\lambda \phi)$& $-5.4$& $-0.54$& $0$&
$-29$& $-0.097$& $0.20$& $0.17$ & $-36$ \\
& $-5.4$& $-0.054$& $310$& $0$& $-0.0017$&
$-0.029$& $0.0014$ & $-1.5$ \\
& $-4.5$& $-0.6$& $0$& $-24$& $-0.082$&
$0.17$& $0.16$ & $-26$\\
& $-34$& $-0.6$& $1900$& $0$& $-0.011$&
$-0.18$& $0.012$ & $-59$\\ \hline
\end{tabular}
\end{center}
\end{table}

In Table \ref{tab:alpha}, we display specific results for both forms of the potential,
$V(\phi)$, considered; both with $\lambda = 5$. We also show results for
both sets of normalization \cite{Webb,Petitjean} and for both a
linear ($\zeta \ne 0$) and quadratic ($\xi \ne 0$) coupling of $\phi$ to $F^2$.
Here, we will not consider the constraints on $\alpha$ when relations between 
all three gauge couplings are assumed \cite{co,other,morebbn}. The latter are typically
a factor of 100 times stronger.  The Oklo bound \cite{Oklo} is roughly
$\Delta \alpha / \alpha \la 10^{-7}$ Therefore any result in Table \ref{tab:alpha}
with a value of $({\Delta \alpha \over \alpha})_{0.14}$ in excess of 0.1 is excluded.
For the $\exp(\lambda\phi^2/2)$ potential, the linear coupling with the $-0.54 \times 10^{-5}$
normalization is excluded, while the other cases are all allowed.  From Fig. \ref{alphaV1},
we see also that smaller values of $\lambda$ lead to larger
variations in $\alpha$ at $z = 0.14$.  For the $\cosh(\lambda\phi)$ potential, all cases
with $\lambda = 5$ are allowed. However, as one can see from Fig. \ref{alphaV2}, 
even the slightly lower values of $\lambda$ shown lead to excessive variations in 
$\alpha$ for the linear coupling.

A closer look at Table \ref{tab:alpha} reveals an interesting observation. A non-zero value for 
$\daa$ at the level of $\sim 0.5 \times 10^{-5}$ at $z = 3$ can be perfectly consistent 
with the bound $|\daa(z=1.5)|\leq 0.6 \times 10^{-6}$ because of the highly non-linear
evolution of the scalar field. Indeed, the quadratic coupling for the cosh-like potential 
shows that the ratio $ \Delta\alpha(z=1.5)/\Delta\alpha(z=3)$ can be as small as 0.01. This implies 
that in certain models the non-zero result of Ref. \cite{Webb} for $\daa$ that spans a rather 
large range of redshifts and the better-sensitivity zero-result measurement of $\daa$ 
at one redshift $z=1.5$ may not necessarily be contradictory. 
This would require, however, that the dominant
source of the effect observed in \cite{Webb} is due to their high redshift
absorbers ($z > 1.8$) which also carry the largest uncertainties.
Nevertheless, we see that even in the simple 
models considered here, the change of $\alpha$ in time can be compatible with both claims
without any special fine-tuning.

The meteoritic bound was discussed in detail in \cite{OPQ}, where the bound
$-8 \times 10^{-7}<\Delta\alpha/\alpha<24 \times 10^{-7}$ was derived.
Note that in this case, the bound relates to the time average of $\Delta\alpha/\alpha$
which is the quantity given in Table \ref{tab:alpha}.  Thus values in the table should be restricted 
between -0.8 and 2.4.  All models shown satisfy this constraint.

The combined result of several recent atomic clock measurements limit
the present day rate of change in the fine-structure constant to $\dot \alpha / \alpha
= (- 0.9 \pm 4.2) \times 10^{-15}$ \cite{fischer}. All models considered easily satisfy this bound.
Finally, we compute the differential acceleration of the Earth-Moon system towards the Sun.
The limit on $\Delta g / g$ is $0.9 \times 10^{-12}$ \cite{EP}. As one can see, all of the models considered
easily satisfy this bound as well.

\subsection{Early time evolution}

The experimental searches for variations in  $\alpha$ 
provide a very useful constraint (or determination, according to Ref. \cite{Webb}) 
on the size of the quadratic term, $\xi= \lambda n/(1+b)$, where we dropped 
the subscripts in $n_F$ and $b_F$. Since $\xi$ is typically $10^{-4}$ or smaller, 
one cannot determine whether this the result should be interpreted as 
$b\gg 1$ or $n\ll 1$ or both. 

However, there are two additional important observables, $\Delta\alpha$ at the cosmic 
microwave background epoch \cite{cmb}, and at BBN \cite{co,morebbn,Noll} 
that will provide more information 
on $b$ and $n$. In fact, since the evolution of $\alpha$ at early times is monotonic, 
and since BBN and CMB data provide a similar level of sensitivity of order a few per
cent to $\daa$, we will just use the BBN constraint as it extends much farther back in time.

 For both potentials, (\ref{Vphi}) and 
(\ref{Vphi1}), we calculate the corresponding values of the field at the BBN redshift, $z\simeq 10^{10}$. 
\begin{eqnarray}
\phi_{BBN} \simeq -5.52 {\rm ~~for~~} V(\phi)= V_0\exp(\lambda\phi^2/2)\\
\phi_{BBN} \simeq -11.9 {\rm ~~for~~} V(\phi)= V_0\cosh(\lambda\phi)
\end{eqnarray}
These values of $\phi_{BBN}$ are calculated for the choice of $\lambda = 5$, 
and can be used to estimate the size of the correction to the fine structure constant,
$\daa_{BBN}$. Alternatively, we 
can use the scaling of the $V(\phi)$ with $a$ in Eq. (\ref{vofphi}) to get the same result. 
We conservatively assume that this quantity is limited at the level of 
$0.06$ (see {\em e.g.} for the latest re-analysis of this bound \cite{cfo5})
based on the new derived $^4$He abundance of $Y_p = 0.249 \pm 0.009$ \cite{os2}.
The connection between $V$ and $B_F$ that we impose (\ref{BF2}) 
does not allow for a linearized form of 
$B_F(\phi)$ (\ref{dp}) at the high redshift associated with BBN, because
the potentials we consider in this paper are very steep.

\begin{center}
\begin{figure}
\vspace{1cm}
\epsfxsize=5.5cm \centerline{\psfig{file=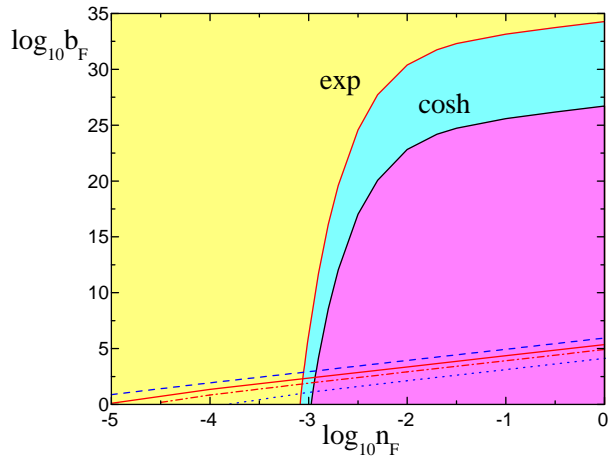, height=6cm}}
\vspace{-1cm}
\caption{ Constraints on the parametrized relation between $B$ and $V$
coming from the early- and late-time evolution of the scalar field.
The constraints from the early-time evolution determine the allowed region in the $b-n$ plane
for the both potentials (light shaded) and for the $\cosh(\phi)$ potential alone (medium shaded).
Both potentials fail the constraint in the dark shaded region.
The early-time evolution constraint is shown by the near horizontal lines at low $b$.
The values for $\xi$ are taken from the table: $\xi \times 10^6 = 5.8, 22, 310$, and 1900 are shown
by the dashed, solid, dot-dashed, and dotted  lines respectively.
} \label{lognbc}
\end{figure}
\end{center}

Using Eq. (\ref{BF}) together with the ansatz (\ref{BF2}), the constraint
$\Delta \alpha / \alpha (\phi_{BBN})$ becomes
\be
\left( {1 + b \over b + V(\phi_{BBN})/V_0} \right)^n > 0.94
\ee
This constraint is shown in Fig. \ref{lognbc} for our two choices of potentials,
both with $\lambda = 5$.
In the light shaded region, both potentials satisfy the BBN bound on $\alpha$.
In the medium shaded region, only the $\cosh(\phi)$ potential satisfies the bound.
In the dark shaded region, both potentials lead to variations in $\alpha$ at the time
of BBN which are in excess of the constraint.
Also shown in Fig. \ref{lognbc} is the constraint obtained in the previous section 
from the late-time evolution of $\phi$.  Here we show the relation
$\xi = \lambda n/ (1 + b)$ for the four values of $\xi$ given in Table 1.
The values of $\xi \times 10^6 = 5.8, 22, 310$, and 1900 are shown
by the dashed, solid, dot-dashed, and dotted  lines respectively.
As one can see, when $n \la 10^{-3}$, the BBN constraint effectively disappears.
At the same time, the choice of $\xi \sim 10^{-4}$ and $n > 10^{-3}$ is clearly 
forbidden by the BBN constraint. Given this constraint on $n$, we can see that only 
very shallow exponential profiles of $B_F(\phi)$ are allowed. 

\section{Coupled quintessence}
\setcounter{equation}{0}

In this section, we investigate changes to the cosmological evolution 
of the scalar field that can be induced by a non-trivial function $B_m(\phi)$. 
The case of the so-called ``coupled quintessence" has been addressed 
in the  literature already \cite{Amendola}, but was specialized to the 
particular case of an exponential potential with a  linear coupling to 
matter. 

The function $B_m(\phi)$ expresses the $\phi$ dependence of the mass of 
any non-relativistic particle, including 
dark matter particles. It is clear that if this mass experiences 
significant changes during the evolution of the Universe,
the scaling of the matter energy density 
will differ from $a^{-3}$, while the number density 
of dark matter particles $n$ will continue to scale as $a^{-3}$. 
The cosmological evolution of 
$\phi$ also changes, because $\phi$ is driven by an effective 
potential that now consists of two parts,
\be
V(\phi)_{\rm eff} = V(\phi) + \rho_m(\phi)=V(\phi)+B_m(\phi)n.
\label{Veff}
\ee
It is not possible to say {\em a priori} which term is more 
important for the cosmological evolution of $\phi$. 

If $B_m(\phi)$ coincides with $B_F(\phi)$, we can use the insight gained
from the previous section to conclude that $B_m(\phi)$ would provide 
no significant change on the cosmology from $z=0$ to $z_{CMB}$. 
This is because the $\daa$ constraints can be re-interpreted as a 
very tight bounds  on $n_F$ and $b_F$, and as a consequence on 
$n_m$ and $b_m$. However, $B_m(\phi)$ and $B_F(\phi)$ need not be the same. 
In fact, in simplest models (for example in the  Brans-Dicke model, or string-inspired
models with a modulus field $\phi$), these functions typically are different. 
In the rest of this section, we assume that $B_m(\phi)$ and $B_F(\phi)$ and not related and 
investigate possible constraints on $n_m$ and $b_m$. 

For definiteness we choose $b_m=1$, and then $B_m(\phi) = (V(\phi))^{n_m}$, according to our 
parametrization. If we choose $n_m > 0$, the effect of $B_m(\phi)$ causes
$\rho_m$ scale faster than $a^{-3}$. In fact, if the dynamics of the scalar field is driven mostly by 
$V(\phi)$ in the radiation dominated epoch, we can derive a simple upper bound on $n_m$. 
Indeed, according to (\ref{vofphi}), the scaling of $V(\phi)$ in the radiation-dominated regime 
is $a^{-4}$. Then it is clear that $B_m(\phi) \sim V(\phi)^{1/4}$ would lead to 
$B_m $  scaling as  $a^{-1}$, 
which also means that $\rho_m \sim a^{-4}$. Thus, $n_m < 1/4$, since for 
$n_m\geq 1/4$ the dark matter energy density would 
be redshifted in the same way or even {\em faster} than radiation. 
This would for example, greatly
increase the energy density of dark matter during BBN. 
Therefore, $n_m \leq 0.25$. 
In fact if we
argue that the energy density in matter should not
contribute more than the energy density in radiation at BBN, we can obtain a 
slightly stronger bound.  For simplicity, let us take the ratio of matter to radiation, $r$, 
today to be $10^4$ (a more accurate number would be $\sim 5500$). 
Then at the time of BBN, the ratio would be
\be
r_{BBN} \sim r_0 a_{BBN}^{(1-4n)}
\ee
For $r_{BBN} \la 1$,  $r_0 \sim 10^4$, and $a_{BBN} \sim 10^{-10}$, we have $n \la 6/40$.

\begin{center}
\begin{figure}
\vspace{1cm}
\epsfxsize=5.5cm \centerline{\psfig{file=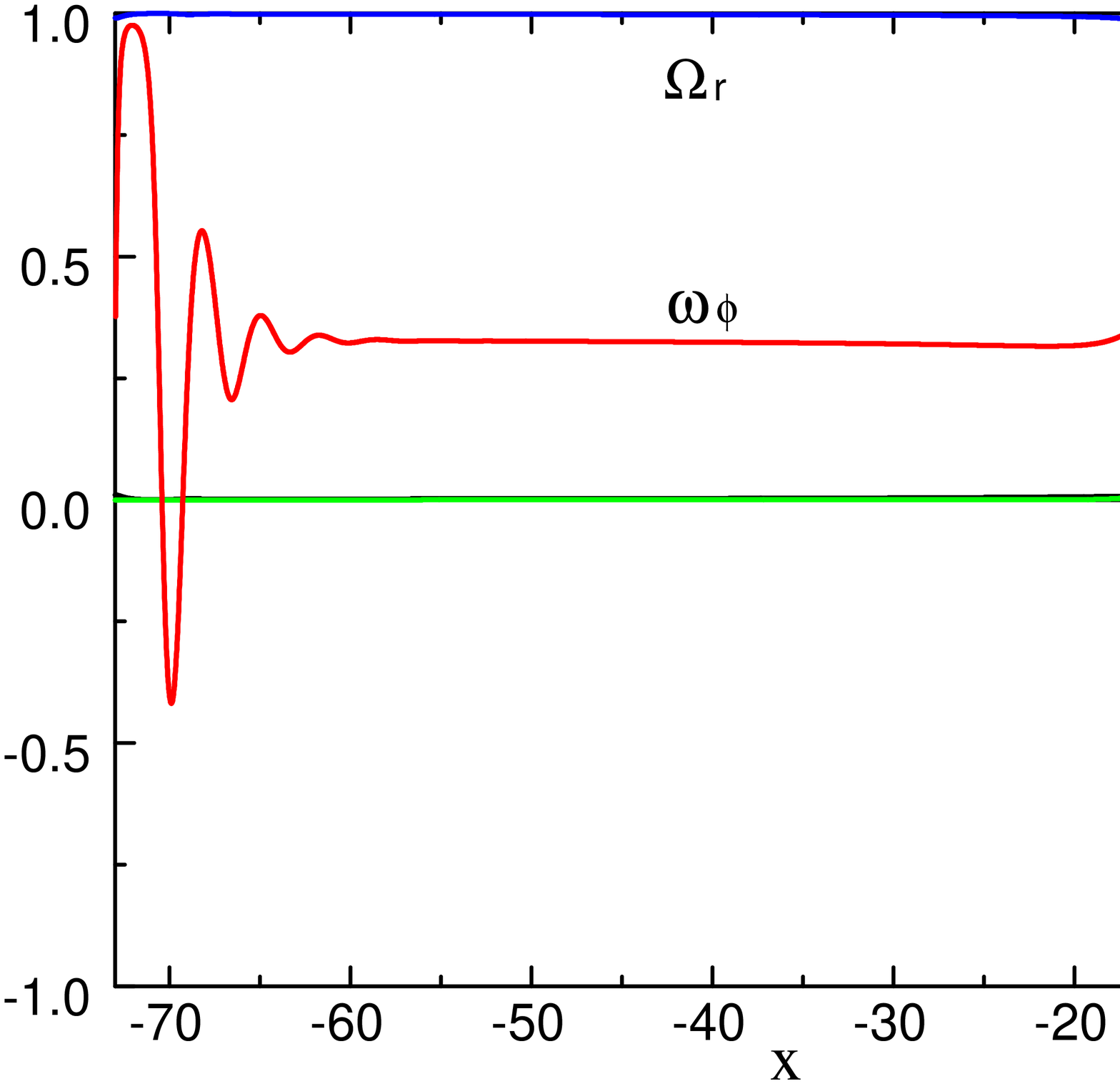, height=6cm}\psfig{file=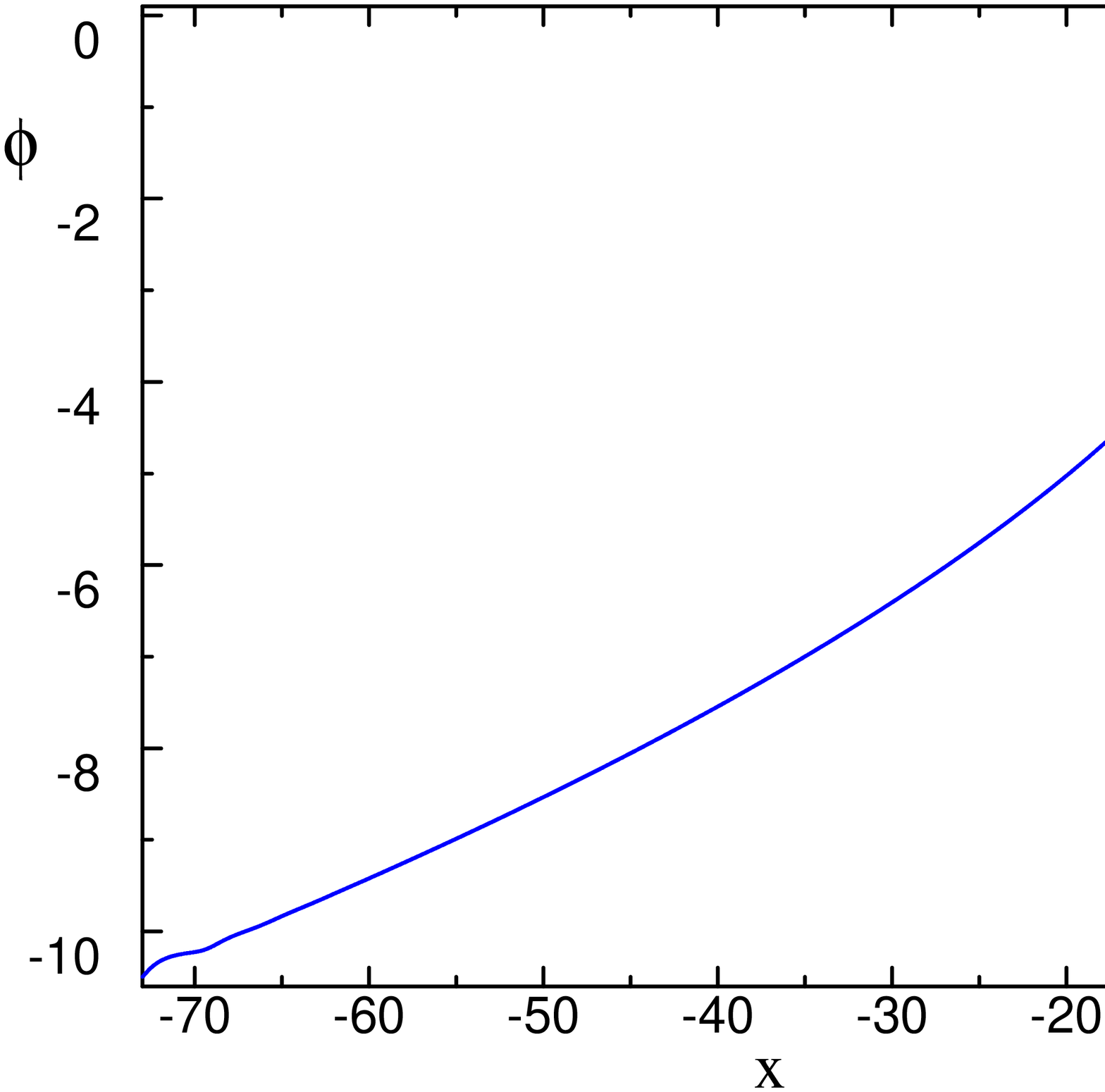, height=6cm}}
\vspace{-1cm}
\caption{ a) The evolution of different components of the energy density as a function of $x$
for the case of $n_m = 0.1$ and $b_m = 1$,  b) The evolution of $\phi$ under the same conditions. 
} \label{lastfig}
\end{figure}
\end{center}

In Fig. \ref{lastfig} we show the changes in the cosmological evolution of the scalar field, matter 
and radiation that result from choosing $n_m = 0.1$. As one can see, there are several important differences from the uncoupled case. 
Because of the change in the mass of the dark matter (as described above),
matter domination occurs somewhat earlier.  In this case, matter domination starts at
$x \simeq - 9.6$ or at $z \simeq 15000$ and ends at $x \simeq - 0.33$ corresponding
to $z \simeq 0.4$.  
Perhaps the most dramatic 
difference is seen in the evolution of the
EOS parameter $\omega_\phi$. Instead of falling from the tracking value of 1/3 towards -1, as
in the uncoupled case, here we see that $\omega_\phi$ increases towards +1, before ultimately
dropping to $-1$ at later times.  This is due to the fact that when the matter density
increases, the effect on the potential $V(\phi)_{\rm eff}$ given in Eq. (\ref{Veff}) drives the
evolution of $\phi$ faster. The field begins to quickly roll and the energy density in $\phi$
becomes dominated by kinetic rather than potential energy. 
The evolution of $\phi$ is shown in Fig. \ref{lastfig}b. As one can see,
around $x \sim -10$, $\phi$ begins to change rapidly.  Hence $\omega_\phi \approx 1$.
(This is also the reason for the curious bump in $\Omega_\phi$ seen in Fig. \ref{lastfig}a.)
However,  ultimately $\phi$ approaches and settles at the minimum of the potential.
At that time, $\omega_\phi$ quickly moves towards -1 and the Universe becomes
dominated by the dark energy.

\begin{center}
\begin{figure}
\epsfxsize=5.5cm \centerline{\psfig{file=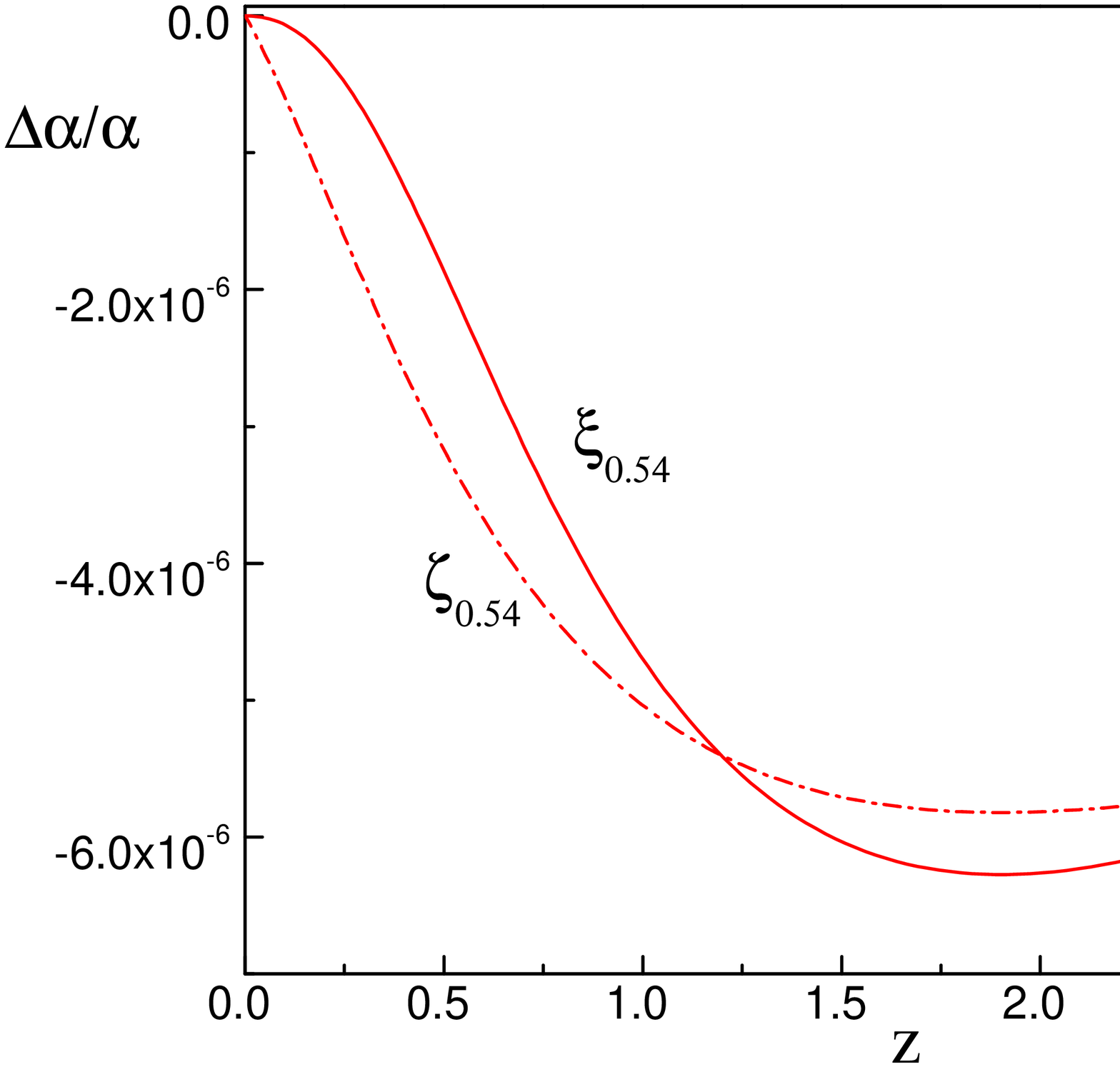, height=6cm}\psfig{file=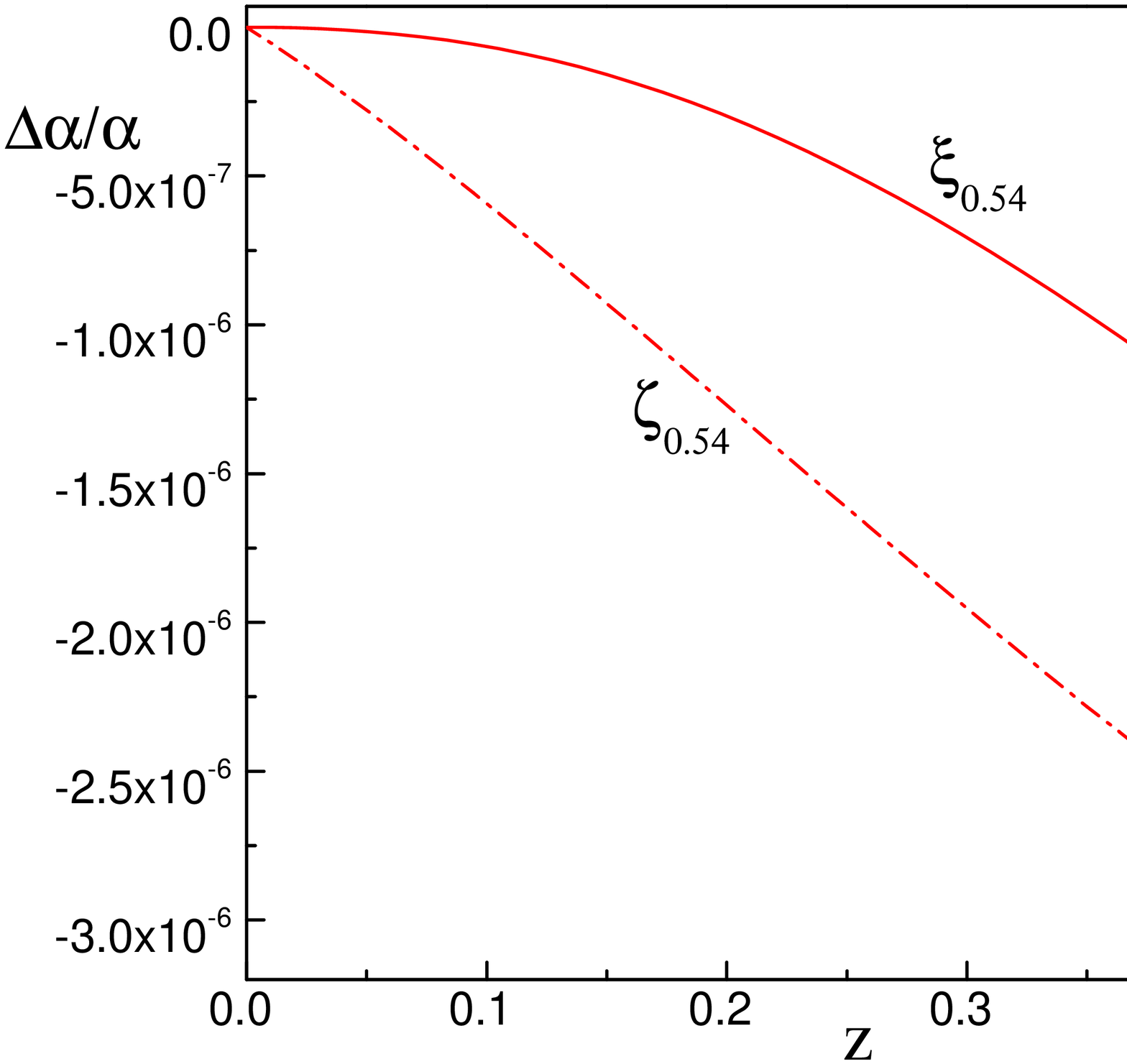, height=6cm}}
\epsfxsize=5.5cm \centerline{\psfig{file=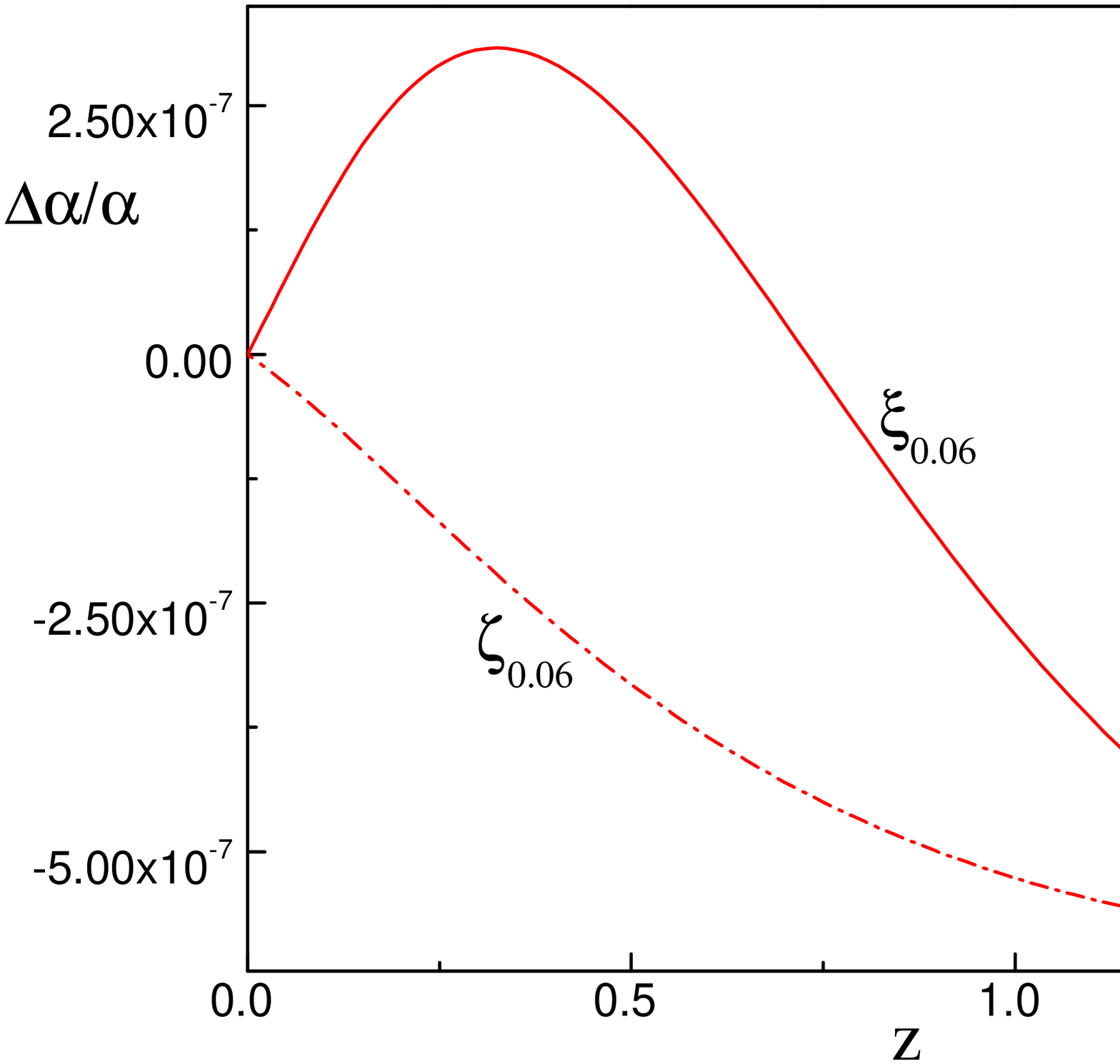, height=6cm}\psfig{file=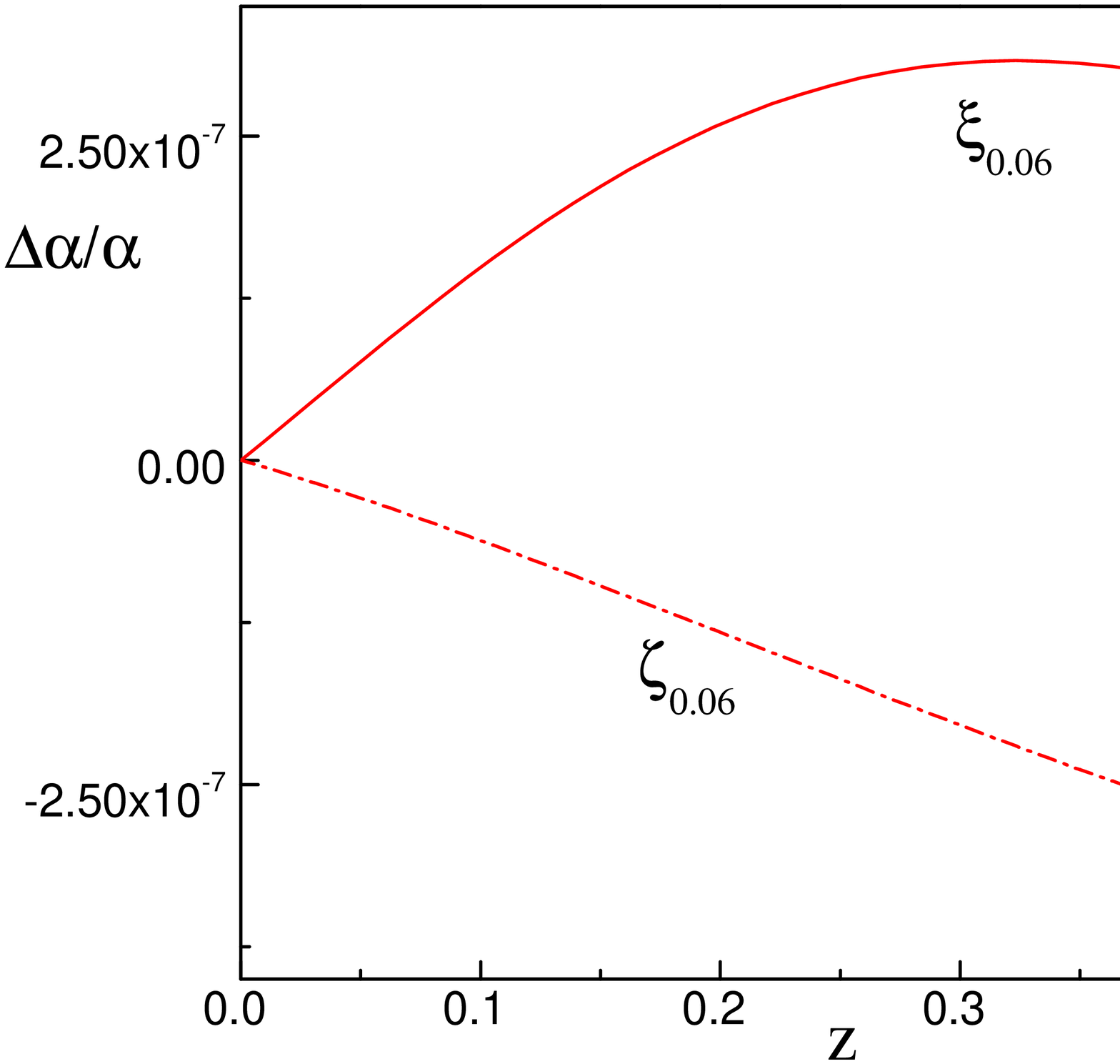, height=6cm}}
\vspace{-1cm}
\caption{ As in Fig. \protect\ref{alphaV1} for the potential (\protect\ref{Vphi}) with $\lambda = 5$.
In this case, $B_m \ne 0$, and we have chosen $n_m = 0.1$ and $b_m = 1$.
} \label{alphaV1c}
\end{figure}
\end{center}

Finally we comment on the effect of the coupling to matter on the variation of $\alpha$.
The evolution of $\Delta \alpha$ is shown in Fig. \ref{alphaV1c} using the potential
(\ref{Vphi}) with $\lambda = 5$.
One should note that because of the matter coupling, the field $\phi$ has moved
quickly and at the present time has already passed the minimum twice
for our choice of parameters. At late times ($z < 3$), the field is significantly closer to
the minimum relative to the uncoupled case. As a result, the normalization
using the QSO measurements results in significantly larger couplings, $\zeta$ and/or $\xi$.
As one can see from the figure and Table \ref{tab:alphac}, despite the fact that the
field is closer to the minimum, the constraints from Oklo and the meteoritic abundances
are in fact more severe. Only the results based on the normalization using
$\daa = 0.06 \times 10^{-5}$ are compatible with these constraints. Similarly 
the differential acceleration is predicted to be much larger in this case, but
the equivalence principle constraints are still satisfied. In contrast,
because the field has nearly settled to its minimum, $\dot \phi$ and hence $\dot \alpha$
is much smaller than in the uncoupled case.

One can also determine the bounds on $n_F$ and $b_F$ for this model.
The late-time evolution constraint is very similar to, but slightly weaker than the result shown in
Fig. \ref{lognbc} for the $\exp(\lambda\phi^2/2)$ case.
The early-time evolution constraint is weaker than all of the cases shown in Fig. \ref{lognbc}
and does not exclude any region of the $b-n$ plane which is not already excluded by the
late-time constraint.

\begin{table}[htb]
\begin{center}
\caption{As in Table \ref{tab:alpha} for the coupled case with $B_m \ne 0$, and
$b_m = 1$ and $n_m = 0.1$.
\label{tab:alphac}}
\vskip .3cm
\begin{tabular}{|c|c|c|c|c|c|c|c|c|}
\hline $\fr{V(\phi)}{V_0}$& $(\fr{\Delta\alpha}{\alpha})_{3}$ &
$(\fr{\Delta\alpha}{\alpha})_{1.5}$ & $\xi$ &$\zeta$&
$(\fr{\Delta\alpha}{\alpha})_{0.14}$ &
$(\fr{\overline{\Delta\alpha}}{\alpha})_{0.45}$ &
$\fr{\dot{\alpha}}{\alpha}$ (yr$^{-1}$) &
$\fr{\Delta g}{g}$ \\
\hline $\exp(\lambda \phi^2/2)$& $-5.4$ & $-5.7$ & $0$&
$-150$& $-0.85$& $-1.3$& $0.38$ & $-1000$ \\
&$-5.4$& $-7.4$& $40000$& $0$& $2.5$&
$2.6$& $-0.034$ & $-13000$ \\
&$-0.56$& $-0.6$& $0$& $-16$& $-0.089$&
$-0.14$& $0.039$ & $-11$ \\
&$-0.44$& $-0.6$& $3200$& $0$& $0.20$&
$0.21$& $-0.0027$ & $-88$ \\
\hline
\end{tabular}
\end{center}
\end{table}

\section{Conclusions}

We have analyzed the cosmological evolution of the scalar field driven by its 
self-interaction potential, $V(\phi)$, and its possible couplings to matter, $B_m(\phi)$. 
Following an original
Damour-Polyakov hypothesis, we assumed a common extremum for  the functions
$V(\phi)$ and $B_m(\phi)$ so that the evolution of the scalar field is driven 
towards that extremum, chosen here to be at $\phi = 0$. With our choices of potentials, 
the evolution of the scalar field occurred in the 
tracking regime throughout the radiation dominated epoch. 
For definiteness we chose two potentials that ensure 
tracking and have a minimum, $V(\phi) = V_0\exp(\lambda\phi^2/2)$ and 
$V(\phi)= \cosh(\lambda\phi)$. 

We have seen that the coupling of the scalar field to the electromagnetic field
is severely restricted by the late-time evolution of the field
due to limits obtained from spectral measurements of quasar absorption systems, the
Oklo phenomenon and meteoritic data. Models with quadratic couplings of the 
scalar field to radiation are typically more sensitive to the quasar absorption spectra-
derived constraints than to the
Oklo bound. The late-time evolution of the scalar field can be very nonlinear, and $\daa$ 
can naturally be close to zero at certain redshifts, while having significant deviations 
over a range of the redshifts.
Thus the experimental results, Refs. \cite{Webb} and \cite{Petitjean}, 
claiming a non-zero result over a range of redshifts $0.5-3$ and zero result at $z=1.5$,
do not have to be viewed as contradictory. 
We also find that the early-time evolution of the field severely constrains the form of the coupling
$B_F(\phi)$ with respect to its relation to $V(\phi)$, due to BBN constraints on 
$\Delta \alpha / \alpha$. The constraints coming from BBN are complementary to those 
from the late-time change in $\alpha$. 

Finally we have considered the case where the quintessence field is coupled
to matter.  Once again, BBN restricts the relationship between $B$ and $V$ due to the
altered scaling of the matter energy density with the scale factor. 
Unlike the uncoupled case, the Universe briefly enters a period where
the equation of state for quintessence is very stiff, before it evolves to a more standard 
dark energy EOS with $\omega_\phi \approx -1$. In order to have a $O(10^{-5}-10^{-6})$ relative 
change of $\alpha$ at redshifts $z\sim 1$, these models typically require larger 
values of coupling between the scalar field and electromagnetic radiation.

\end{document}